\documentclass[aps,prb,twocolumn,superscriptaddress,showpacs]{revtex4-1}
\pdfoutput=1
\usepackage{graphics}
\usepackage{subfigure}
\usepackage{longtable}
\usepackage{mathptmx}
\usepackage{graphicx}
\usepackage{pstricks}
\usepackage{amsmath}
\usepackage{dcolumn}
\usepackage{bm}

\newcommand{\mbf}{\mathbf} 
\renewcommand{\k}{{\mbf k}}

\newcommand{\q}{{\mbf q}}
\newcommand{\filla}{6.12}
\newcommand{\fillb}{6.25}
\newcommand{\Eprint}[2]{\href{#1}{#2}}

\begin{document}

\title{Spin fluctuations and superconductivity in K$_x$Fe$_{2-y}$Se$_2$}

\author{A. Kreisel}
\affiliation{Department of Physics, University of Florida,
Gainesville, Florida 32611, USA}
\author{Y. Wang}
\affiliation{Department of Physics, University of Florida,
Gainesville, Florida 32611, USA}
\author{T. A. Maier}
\affiliation{Center for Nanophase Materials Sciences and Computer
Science and Mathematics Division, Oak Ridge National Laboratory,
Oak Ridge, Tennessee 37831-6494, USA}
\author{P. J. Hirschfeld}
\affiliation{Department of Physics, University of Florida,
Gainesville, Florida 32611, USA}
\author{D. J. Scalapino}
\affiliation{Department of Physics,
University of California, Santa Barbara, California 93106-9530, USA}

\date{August 25, 2013}

\begin{abstract}  Superconductivity in alkali-intercalated iron selenide, with $T_c$'s of $30\,\text{K}$ and above, may  have a different origin than that of the other Fe-based superconductors, since it appears that
  the Fermi surface does not have any holelike sheets centered around the $\Gamma$ point. Here we investigate the symmetry of the superconducting gap in the framework of spin-fluctuation pairing calculations using  density functional theory bands downfolded onto a three-dimensional (3D), ten-orbital tight-binding model,  treating the interactions in the random-phase approximation (RPA). We find a leading instability towards a state with $d$-wave symmetry, but show that the details of the gap function depend sensitively on  electronic structure.  As required by crystal symmetry, quasi-nodes on electron pockets always occur, but are shown to be either horizontal, looplike or vertical depending on details. A variety of other 3D gap structures, including  bonding-antibonding $s$-symmetry states which change sign between inner and outer electron pockets are found to be subdominant. We then investigate the possibility that spin-orbit coupling effects on the one-electron band structure, which lead to enhanced splitting of the two $M$-centered electron pockets in the 2-Fe zone, may stabilize the bonding-antibonding $s_\pm$-wave states.  Finally, we discuss our results in the context of current phenomenological theories and experiments.
\end{abstract}


\maketitle

\section{Introduction}
\label{sec_1}
Recent investigations of the alkali iron selenide materials with nominal composition $A$Fe$_2$Se$_2$ (A-122) have revealed significant differences relative to the other iron chalcogenide and iron pnictide superconductors\cite{stewart_review,HKM11,dagotto_review,hhwen_review}. While most of the materials in the class of iron-based superconductors (FeSC) show Fermi surfaces with both hole-like and electron-like sheets, angle resolved photoemission (ARPES) studies of the superconducting A-122 systems suggest the absence of the hole-like Fermi sheets around the $\Gamma$ point in the first Brillouin zone\cite{Qian11}. Intriguingly, similar ARPES results have been reported on high-$T_c$ superconducting monolayer FeSe films.  In both cases, this suggests  that the usual arguments leading to $s_\pm$ pairing by repulsive interactions acting between hole and electron pockets must break down.

Various authors have already discussed this problem.
Two dimensional (2D) fluctuation-exchange calculations\cite{Wang11,Maier11,Maiti} using a tight-binding model of the electronic structure in the 1-Fe Brillouin zone found a nodeless $d$-wave state to be the leading instability. It was pointed out, however, that the body-centered tetragonal symmetry of the 122 structure forced a hybridization of the electron pockets, leading to nodes in the $d$-wave gap function on these sheets\cite{Mazin11,Khodas12}.  Saito \textit{et al.} \cite{Kontani11} found such nodal $d$-wave states in 2D random-phase approximation (RPA) spin-fluctuation calculations for a ten-orbital Hubbard model for planes with a few fixed $k_z$
to be subdominant compared to $s_{++}$ states when considering phonon-mediated electron-electron interactions. Mazin, as well as Khodas and Chubukov, pointed out that the hybridization forced by the 122 crystal structure can lead to the possibility of a novel ``bonding-antibonding'' $s_\pm$-wave state with opposite signs on the two $M$-centered hybridized electron pockets in the 2-Fe zone\cite{Mazin11,Khodas12}. There have also been proposals for $s$-wave states with like signs on all of the electron sheets.  For example, Wang \textit{et al.}\cite{Wang11} used a functional renormalization group approach in which a type $s_\pm$ pairing   arose from scattering between states
belonging to a $\Gamma$-centered hole band located below the Fermi level to the electron pockets. Fang \textit{et al.}\cite{Fang11} discussed a strong coupling approximation in which a next-nearest-neighbor exchange $J_2$ along with a Gutzwiller band narrowing leads to an $s_{++}$ state.

Thus proposals for the pairing mechanism of the $A$Fe$_2$Se$_2$ superconductors  range from weak coupling fluctuation exchange, to  superexchange via selenium $4p$ orbitals, to orbital correlation enhanced phonon exchange scenarios. The different results for the gap structure predicted by these  theories should provide an opportunity to differentiate between them experimentally. Here we will focus on the fluctuation-exchange mechanism with the goal of determining the consequences of the absence of the hole-like Fermi sheets and the resulting structure of the gap. In these calculations we will treat the full 3D system and explore the differences with arguments based on simpler 2D models.

It is important to recall that the exact crystal structure of the optimal superconducting material of this type has not yet been determined. Current superconducting samples may be inhomogeneous at the nanoscale or mesoscale, consisting of regions of ordered magnetism and vacancies\cite{WBao_2011} and superconducting
regions with fewer, possibly disordered vacancies.  Some aspects of the inhomogeneity have been reviewed in Refs. \onlinecite{HKM11}-\onlinecite{hhwen_review}.  We attempt here to describe the effective electronic structure of the superconducting regions with a model similar to the doped $A_x$Fe$_{2-y}$Se$_2$
system.   A similar effective band  model may also arise from scattering from disorder in the $\sqrt{5}\times\sqrt{5}$
Fe vacancy reconstructed lattice\cite{Berlijn_2012}.  Ultimately, we believe  that the general types of order parameter structures which emerge from our calculation should be at least qualitatively correct if the effective bands are similar to those
observed in ARPES,  and if the pairing vertex can be described
by an RPA spin-fluctuation interaction.

This paper is organized as follows.
In Sec. \ref{sec:Model} we introduce a 3D ten-orbital Hubbard-Hund model which will be used to describe the 2-Fe per unit cell A-122 materials. The band structure and the Fermi surfaces for two representative band fillings are described, and we briefly review   the fluctuation-exchange approximation to the pairing vertex    and the linearized gap equation.  We discuss how the 3D 10 Fe-orbital tight-binding Hamiltonian adopted can capture certain symmetry effects which are missing in the 2D descriptions.
We next present in Sec.~\ref{sec:Results} our results for the pairing vertices and gap functions, and show that for the given nonrelativistic density functional theory (DFT)-based tight-binding bands, the leading states are always $d_{x^2-y^2}$ states with quasi-nodes on the \textit{M}-centered pockets not enforced by $d$-wave symmetry.  Other $d$-wave states and
bonding-antibonding $s_\pm$-wave states are the next leading pairing instabilities, but are found to be strongly suppressed in this treatment.
In Sec.~\ref{sec:hybrid}, we consider physical effects, including spin-orbit coupling, neglected in the ``standard" approach that can hybridize the electron pockets significantly and tend to stabilize the bonding-antibonding $s_\pm$-state. We find however that, within a physical reasonable parameter range, the $d_{x^2-y^2}$ pairing channel remains stable.
In Sec.~\ref{sec:conclusions} we finally present our conclusions.

\section{Model}
\label{sec:Model}
\subsection{Band structure}
\label{subsec:band structure}
\begin{figure}[tb]
 \includegraphics[width=1.0\linewidth]{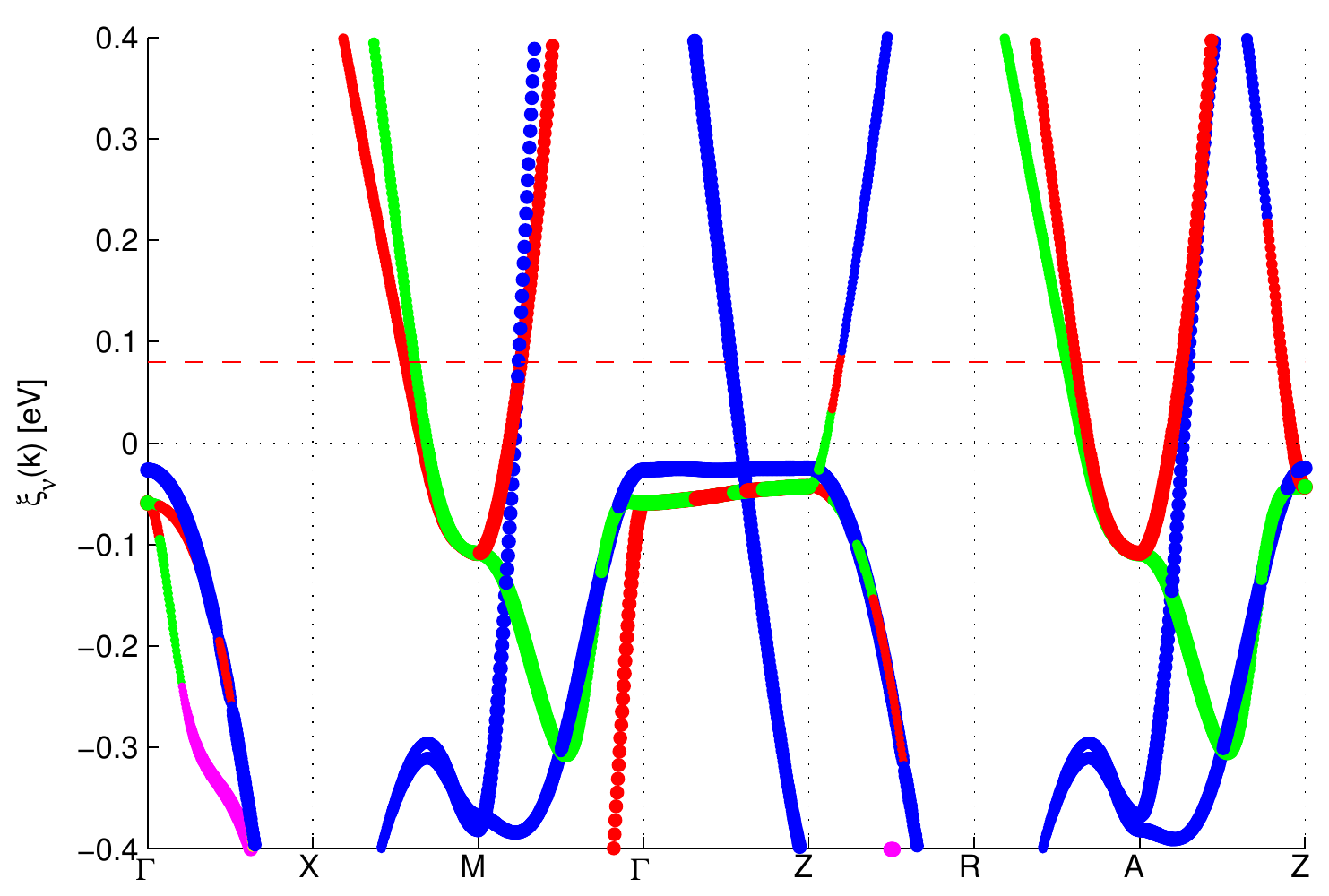}
\caption{(Color online) Details of the bandstructure of the tight-binding model for K$_x$Fe$_{2-y}$Se$_2$ where a shift of the chemical potential has been applied to obtain a filling of $n=\filla$ (underdoped case). The orbital character is indicated by the colors red $d_{xz}$, green $d_{yz}$, blue $d_{xy}$, yellow $d_{x^2-y^2}$, and purple  $d_{3z^2-r^2}$ and the value of the largest orbital weight is proportional to the radius of the dots. The dotted line represents the Fermi level for filling $n=\filla$ while the dashed (red) line is the Fermi level for filling $n=\fillb$ (overdoped case).
 \label{fig0}}
\end{figure}
The nonrelativistic electronic structure for K$_x$Fe$_{2-y}$Se$_2$ has been obtained using the WIEN2k~\cite{Blaha2001} package in conjunction with the WIEN2Wannier addon and the Wannier90 program.
After the projection onto Wannier orbitals, the nearest-neighbor hoppings $t_x(d_{xz},d_{xz})$ and $t_x(d_{xy},d_{xy})$ and the symmetry related hoppings (labeled in the 1 Fe description) were reduced by $\delta t=0.0625\,\mathrm{eV}$ to approximately match the features of the Fermi surface seen by ARPES experiments\cite{Qian11,Chen11}. Details of the band structure described by the Hamiltonian
\begin{equation}
 H_0=\sum_\sigma \sum_{ij} \sum_{\ell \ell'} t_{ij}^{\ell \ell'} c_{i\ell\sigma}^\dagger c_{j\ell' \sigma}\,,\label{eq_H0}
\end{equation}
where $t_{ij}^{\ell \ell'}$ are the hoppings connecting sites $i$ and $j$ for orbitals $\ell$ and $\ell'$, are shown Fig.~\ref{fig0} for energies close to the Fermi level.

\begin{figure*}[tb]
 \includegraphics[width=\linewidth]{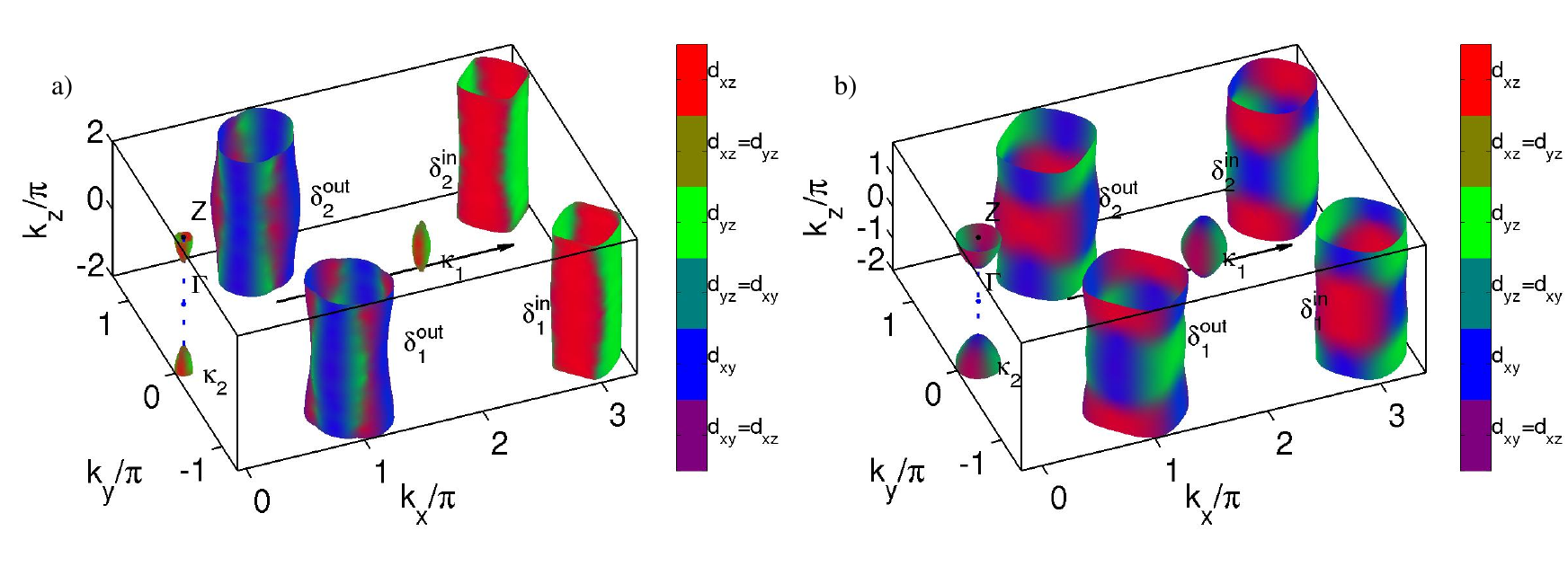}
\caption{(Color online) Orbital character of the Fermi surface shown in $k$-space (set the lattice constants $a=c=1$) plotted in the conventional (2-Fe)  reciprocal unit cell of KFe$_2$Se$_2$ for filling (a) $n_1=\filla$ and (b) $n_2=\fillb$ , red $d_{xz}$, green $d_{yz}$, and blue $d_{xy}$ visualized with the summed-color method where the absolute value of the overlap is mapped to the RGB value of the color on the surface. Note that the inner electron pockets have been shifted artificially by $2\pi$ in the $k_x$ direction to make them visible; the shift vector is indicated by a black arrow.
 \label{fig1}}
\end{figure*}

Within the approximation to the electronic pairing vertex described below,
the electronic structure enters the pairing strength eigenvalue problem Eq.~(\ref{eqn:gapeqn}) only through the Fermi surface.
Due to the sample preparation issues alluded to above, several possible dopings have been
reported for superconducting samples.
In this investigation, we consider two doping levels, (see Fig. 1), which represent different ends of the superconducting range in order to illustrate the effect of doping on the pairing.
Although there is no clear empirical relationship between $T_c$ and doping in
this material, we will refer to the doping with $n=\filla$ electrons as ``underdoped";
it is close to the critical doping where the $\Gamma$-centered hole pocket disappears\cite{Maier12}, and corresponds nominally\cite{Note1} to the alloy
   K$_{0.8}$Fe$_{1.7}$Se$_2$.  We also consider an ``overdoped" filling $n=\fillb$, which corresponds to K$_{0.85}$Fe$_{1.8}$Se$_2$.

The Fermi surfaces derived from our tight-binding model  shown in Fig. \ref{fig1} exhibit two significant differences between the two dopings: In the underdoped case,  electron like pockets centered at the $Z$ point  (labeled by $\kappa_1,\kappa_2$) are negligibly small,
suggesting that 2D calculations performed previously\cite{Maier11,Wang11} may be accurate,
with the exception of features attributable to the peculiar 122 crystal symmetry\cite{Mazin11}.
In the overdoped case $n=\fillb$, on the other hand, these $Z$-centered pockets make a significant contribution
to the Fermi level density of states. Note that the orbital weight on the pointed oval shaped $Z$ pockets is mainly $d_{xz}$ and $d_{yz}$, whereas near the poles along $\Gamma$-$Z$ and $Z$-$R$ it is $d_{xy}$ (see Figs.~\ref{fig0} and \ref{fig1}). In addition, the dispersion of the hybridized bands that lead to the $M$-centered electron like pockets  makes the orbital weight of the inner and outer  pockets  (labeled by $\delta_1^{\rm in,out},\delta_2^{\rm in,out}$) look quite different between the two doping levels: While in the underdoped case the outer electron-like pocket is mainly $d_{xy}$ in character (apart from some spots close to the hybridization lines) and the inner pocket shows vertical stripes of $d_{xz}$ or $d_{yz}$ character, in the electron doped case the hybridization of the sheets occurs on a horizontal line leading to spots of orbital weight of all three orbitals on the inner and outer sheets.

\subsection{Spin fluctuation pairing}

The local interactions are included via the ten-orbital Hubbard-Hund Hamiltonian
\begin{eqnarray}
	H = H_{0}& + &\bar{U}\sum_{i,\ell}n_{i\ell\uparrow}n_{i\ell\downarrow}+\bar{U}'\sum^\prime_{i,\ell'<\ell}n_{i\ell}n_{i\ell'}
	\nonumber\\
	& + & \bar{J}\sum^\prime_{i,\ell'<\ell}\sum_{\sigma,\sigma'}c_{i\ell\sigma}^{\dagger}c_{i\ell'\sigma'}^{\dagger}c_{i\ell\sigma'}c_{i\ell'\sigma}\\
	& + & \bar{J}'\sum^\prime_{i,\ell'\neq\ell}c_{i\ell\uparrow}^{\dagger}c_{i\ell\downarrow}^{\dagger}c_{i\ell'\downarrow}c_{i\ell'\uparrow} \nonumber \label{H}
\end{eqnarray}
where the interaction parameters $\bar{U}$, $\bar{U}'$, $\bar{J}$, $\bar{J}'$ are given in the notation of Kuroki \textit{et al.} \cite{Kuroki08}. Here, $\ell$ is an orbital index with $\ell\in(1,\ldots,10)$ corresponding to the Fe-orbitals $(d_{xz},d_{yz},d_{xy},d_{x^2-y^2},d_{3z^2-r^2})$ on the first and second iron atom in the elementary cell. Note that the $\sum^\prime$ only gives a contribution when the indices $\ell$ and $\ell'$ label an orbital on the same iron atom.
The corresponding Fermi surface together with the orbital character for the two doping levels with fillings $n_1=\filla$ and $n_2=6.25$ are shown in Fig.~\ref{fig1}.
\label{subsec:spinfluct}
Given the Green's functions in our model with the eigenenergies $\xi_\nu(\k)$ measured from the Fermi level
\begin{equation}
	G^\nu(k)= \frac{1}{i\omega_n - \xi_\nu(\k)},
\end{equation}
where we introduced the 4-momenta via $k=(\k,\omega_n)$, we now calculate the susceptibility in the normal state as
\begin{eqnarray}
	\chi_{\ell_1 \ell_2 \ell_3 \ell_4}^0 (q) & = & - \sum_{k,\mu\nu} M_{\ell_1\ell_2\ell_3\ell_4}^{\mu\nu} (\k,\q)
	 G^{\mu}(k+q) G^{\nu} (k).   \label{eqn_supersuscept}
\end{eqnarray}
Here the matrix elements relating band and orbital space, $a^{\ell}_{\nu}(\k)=\langle d_\ell|\nu k\rangle$ combine to the tensor
\begin{equation}
	M_{\ell_1 \ell_2 \ell_3 \ell_4}^{\mu\nu} (\k,\q) = a_\nu^{\ell_4} (\k) a_\nu^{\ell_2,*} (\k) a_\mu^{\ell_1} (\k+\q) a_\mu^{\ell_3,*} (\k+\q).
\end{equation}
Taking into account the interactions cited in Eq.~(\ref{H}) in a random phase approximation (RPA) framework, we define the spin- ($\chi_1^{\rm RPA}$) and orbital-fluctuation ($\chi_0^{\rm RPA}$) parts of the RPA susceptibility for $q=(\q,\omega_n=0)$
\begin{subequations}
\begin{align}
\label{eqn:RPA} \chi_{1\,\ell_1\ell_2\ell_3\ell_4}^{\rm RPA} (\q) &= \left\{ \chi^0 (q) \left[1 -\bar U^s \chi^0 (q) \right]^{-1} \right\}_{\ell_1\ell_2\ell_3\ell_4},\\
 \chi_{0\,\ell_1\ell_2\ell_3\ell_4}^{\rm RPA} (\q) &= \left\{ \chi^0 (q) \left[1 +\bar U^c \chi^0 (q) \right]^{-1} \right\}_{\ell_1\ell_2\ell_3\ell_4}.
\end{align}
These susceptibilities will be evaluated at low temperatures where they have saturated and no longer change.The spin susceptibility at $\omega=0$ is then given by the sum
\end{subequations}
\begin{equation}
	\label{eqn:chisum} \chi (\q) = \frac 12 \sum_{\ell_1 \ell_2} \chi_{\ell_1 \ell_1 \ell_2\ell_2}^{\rm RPA} (\q)\,.
\end{equation}
The interaction matrices $\bar U^s$ and $\bar U^c$ in orbital space consist of linear combinations of the interaction parameters, and their forms are given e.g.\/ in Ref.~\onlinecite{a_kemper_10}.

Next, we define the scattering vertex in the singlet channel
\begin{eqnarray}
	{\Gamma}_{ij} (\k,\k') & = & \mathrm{Re}\sum_{\ell_1\ell_2\ell_3\ell_4} a_{\nu_i}^{\ell_1,*}(\k) a_{\nu_i}^{\ell_4,*}(-\k) \\
	&&\times \left[{\Gamma}_{\ell_1\ell_2\ell_3\ell_4} (\k,\k') \right] a_{\nu_j}^{\ell_2}(\k') a_{\nu_j}^{\ell_3}(-\k')\,,\nonumber \label{eq:Gam_ij}
\end{eqnarray}
where $\k$ and $\k'$ are quasiparticle momenta restricted to the pockets $\k \in C_i$ and $\k' \in C_j$, where $i$ and $j$ correspond to the band index of the  Fermi surface sheets. The vertex function in orbital space $\Gamma_{\ell_1\ell_2\ell_3\ell_4}$ describes the particle-particle scattering of electrons in orbitals $\ell_2,\ell_3$ into $\ell_1,\ell_4$.
    In RPA it is given by\cite{Takimoto04}:
\begin{eqnarray}
	&&{\Gamma}_{\ell_1\ell_2\ell_3\ell_4} (\k,\k') = \left[\frac{3}{2} \bar U^s \chi_1^{\rm RPA} (\k-\k') \bar U^s\nonumber \right.\,~~~~~~\,\\
	&&\,~~~~~\left. +  \frac{1}{2} \bar U^s - \frac{1}{2}\bar U^c \chi_0^{\rm RPA} (\k-\k') \bar U^c + \frac{1}{2} \bar U^c \right]_{\ell_1\ell_2\ell_3\ell_4}. \label{eq:fullGamma}
\end{eqnarray}

Within the fluctuation-exchange approach, the pairing eigenfunction for a given set of parameters corresponds to the leading eigenvalue of the weighted scattering vertex $\Gamma(\k,\k')$ in the singlet channel. Technically, we can calculate the pairing strength\cite{s_graser_09} $\lambda_\alpha$ for different pairing channels $\alpha$ as eigenvalues of the linearized gap equation
\begin{equation}\label{eqn:lingap1}
  -\frac{1}{V_G}\int_{V_G} d^D \k'\, \Gamma(\k,\k') \delta(\xi_{\k'}) g_\alpha(\k')=\lambda_\alpha g_{\alpha}(\k),
 \end{equation}
where the integral is performed over the $D$-dimensional Brillouin zone with volume $V_G$. In $D=3$ this reduces to
 \begin{equation}\label{eqn:gapeqn}
-\frac{1}{V_G}  \sum_j\int_{\text{FS}_j}dS'\; \Gamma_{ij}(\k,\k') \frac{ g_\alpha(\k')}{|v_{\text{F}j}(\k')|}=\lambda_\alpha g_{\alpha}(\k),
 \end{equation}
 where $v_{\text{F}j}(\k')$ is the Fermi velocity of band $j$ and the integration is performed over the Fermi surface $\text{FS}_j$.
 In this formulation, our one electron basis diagonalizes the bilinear band Hamiltonian and the contribution of interband
pairs ($\k \uparrow \nu, -\k \downarrow \mu$) are neglected for $\mu\ne \nu$.  These processes are cut-off at low energies by $|\xi_{\nu}(\k)-\xi_{\mu}(-\k)|$.
The eigenfunction $g_\alpha(\k)$ for the largest eigenvalue then determines the symmetry and structure of the leading pairing instability and provides an approximate form for the superconducting gap $\Delta(\k)\sim g(\k)$ at least close to $T_c$.
 The eigenvalue may also be expressed directly in
terms of the eigenfunction and pair vertex as
\begin{align}\label{eq:lambda_explicit}
& \lambda [g(k)]=
\notag
 \\
 &- \frac{1}{(2\pi)^4\mathcal N}\sum_{ij} \int_{C_i} \frac{d k_\parallel}{v_\mathrm{F}(k)} \int_{C_j}
\frac{d k_\parallel'}{v_\mathrm{F}(k')} g(k) {\Gamma}_{ij} (k,k')
g(k'),
\end{align}
where $C_i$ is the Fermi surface corresponding to the band $i$, and
\begin{equation}
 \mathcal N=\frac{1}{(2\pi)^2}\sum_i \int_{C_i} \frac{d k_\parallel}{v_\mathrm{F}(k)} [g(k)]^2\label{eq_normal}
\end{equation}
is the normalization that is chosen to be unity within this paper to make the results for the magnitude of the gap function independent of the number of points on the Fermi surface.  We may also decompose the various contributions to the pairing according to Fermi pocket by defining a pocket index $\tau$ by
$\int_{C_i} =\sum_{\tau_i}\int_{C_{\tau_i}}$, where $\tau_i$ runs over all distinct Fermi sheets associated with band $i$.  Then
\begin{equation}
\lambda = \sum_{\tau\eta} \lambda_{\tau\eta}\label{eq:tau_eta}
\end{equation}
sums all contributions involving all pair scattering processes between pocket $\tau$ and $\eta$, but includes all symmetry-equivalent contributions into a given $\lambda_{\tau\eta}$.

{
The Matsubara sum in Eq.~(\ref{eqn_supersuscept}) is carried out in the usual way and we evaluate $\chi_{\ell_1 \ell_2 \ell_3 \ell_4}^0$   by integrating over the full Brillouin zone.
To solve Eq.~(\ref{eqn:gapeqn}), we then use this matrix together with $\bar U^c$ and $\bar U^s$ to construct the pairing vertex ${\Gamma}_{ij} (\k,\k')$.}
Finally, the two-dimensional area of the Fermi surface sheets is discretized using a Delaunay triangulation together with a simplification algorithm that keeps the angle between the surfaces of neighbored triangles small to transform the integral equation Eq. (\ref{eqn:gapeqn}) into an ordinary matrix equation which is solved numerically. Typically we chose a $\k$-mesh of $50\times 50 \times 20$ points for the momentum integration and totally $\approx 1200$ points on all Fermi surfaces to obtain a reasonable convergence both in the bare susceptibility and the pairing calculation.
\begin{figure*}[tb]
 \includegraphics[width=\linewidth]{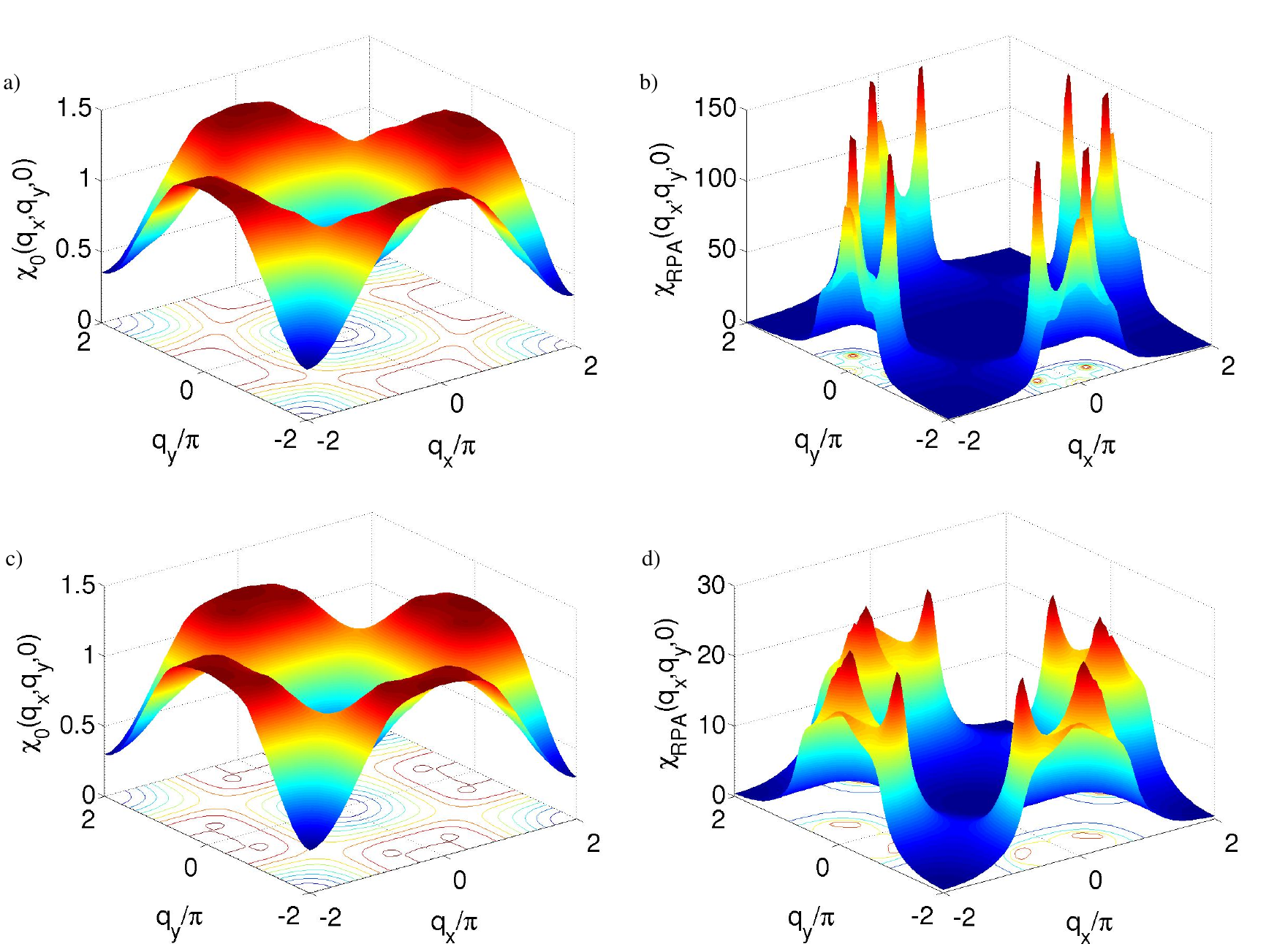}
\caption{(Color online) Bare susceptibility (left) and RPA spin susceptibility (right) using $U=0.88\,\text{eV}$ and $J=U/4$ for the underdoped case $n_1=\filla$ (a), (b) in the $k_z=0$ plane; the same for the overdoped case $n_2=\fillb$ (c), (d). (The susceptibilities at other cuts in $k_z$ as well as for the electron doped case have overall similar properties.)
 \label{fig3}\label{fig_susc}}
 \end{figure*}

\subsection{Symmetry of pair states}
\label{subsec:symmetry}
The early treatments of the pairing problem for $A$Fe$_2$Se$_2$ used simplified 2D models with 1 Fe (five $d$ orbitals) per unit cell, in which case the electron pockets are well separated in momentum space around the $X$ and $Y$ points of the 1-Fe Brillouin zone.  The problem of pairing of electrons in the presence of repulsive interactions   is therefore similar to the study of  Agterberg \textit{et al.}\cite{ABG},  where simplified interactions between and within pockets at high symmetry points were
 studied.  Indeed, the nodeless $d$-wave state obtained in these treatments \cite{Maier11,Wang11}  is analogous to that obtained in 3D in Ref. \onlinecite{ABG} and arises simply from the requirement that in the presence of strong repulsive interpocket interactions, the gap function must change sign;
 $d_{x^2-y^2}$ is favored over $d_{xy}$ because within the 1-Fe unit cell model the nodal lines in the Brillouin zone corresponding to the latter case are located on the Fermi surface, while
 those for the former case are not.     There are
 two essential differences in the realistic problem of pairing in Fe-based superconductors with the 122 structure considered here.  The first one is
   generic to all of the superconductors discovered thus far and is derived from a proper symmetry analysis of the single FeAs or FeSe layer; it involves
   the symmetries of the electronic Bloch states imposed by the out-of-plane As or Se atoms\cite{Lee08,Hu13B,*Hu13a,Vafek13,Fischer13,Sorella13},  and has been shown to lead to the possibility of novel pair states not accessible within 1-Fe  descriptions of the electronic structure.

A second complication arises in the materials with 122 crystal structure, which has $I4/mmm$ space group symmetry rather than the simpler $P4/nmm$ symmetry characteristic of a single FeAs or FeSe layer, or of a crystal consisting of a stack of such layers in which the pnictogen or chalcogen atoms displacements are ``in phase".  In a system with $P4/nmm$ symmetry, the one Fe per unit cell Brillouin zone, with one electron pocket each around the $X$ and $Y$ points
may be ``folded" exactly, yielding  two electron pockets which cross each other at the $M$ points and the $M$-$A$ line in the 2-Fe zone\cite{HKM11}.  In a 122 crystal ($I4/mmm$),
these electron bands hybridize at generic $k_z$ values, splitting the Fermi surface pockets into distinct ``inner" and ``outer" sheets.
This splitting has
consequences for the pairing symmetry in these materials, as pointed out by Mazin\cite{Mazin11}; in particular, the nodeless $d$-wave state
found in the naive 2D five-orbital theories must acquire nodes at these points.  As observed by Maier \textit{et al.}\cite{Maier11}, however, these nodes are not
determined by the $d$-wave symmetry per se, and therefore contribute phase space to physical observables which scales with the hybridization strength, which may be small.  Maier \textit{et al.} referred to these ``narrow" nodes in this limit as ``quasi nodes."  The hybridization  also  allows, in principle, a novel $s$-wave gap\cite{Mazin11} which changes sign between the hybridized electron  pockets around the $M$ point, the so-called ``bonding-antibonding $s_\pm$ state."

Khodas and Chubukov\cite{Khodas12} considered the effect of the hybridization term on the pairing of electrons on the
electron pockets, treating interactions as momentum independent and  the hybridization energy and ellipticity of the electron pockets perturbatively.  They showed that within this simplified model the competition between $s$- and $d$-wave states was controlled by a  parameter $\kappa$ equal to the ratio of the  hybridization to the ellipticity.   In the limit of  small $\kappa$ (small  hybridization), the $d$-wave state was found to be the ground state, while for large $\kappa$ the bonding-antibonding $s$-wave state was favored, and for intermediate values an $s\!+\!id$ state was found.    In a second publication,  these authors considered the detailed nodal structure of the bonding-antibonding $s_\pm$-wave state, showing that depending on details of the Fermi surface,  either vertical line nodes, loop nodes, or a nodeless state could be obtained\cite{Khodas12_PRB}.  In the limit of large hybridization, nodes were argued to disappear.

\section{Results}
\label{sec:Results}

\begin{figure*}[tb]
 \includegraphics[width=\linewidth]{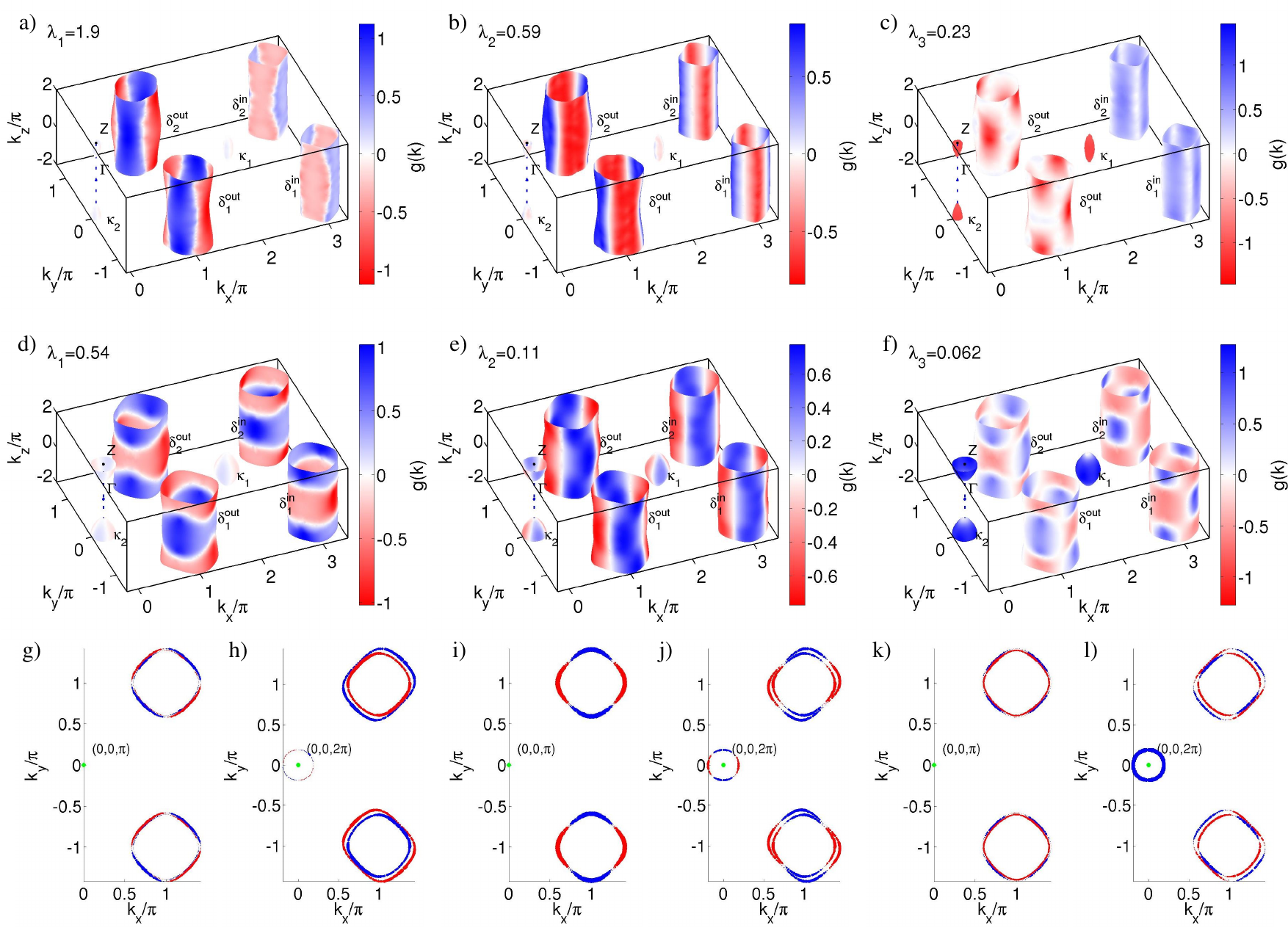}
\caption{(Color online) Gap functions in the 2-Fe Brillouin zone for the leading instabilities obtained using rotational invariant interaction parameters with $U=0.88\;\text{eV}$ and $J=U/4$. Top row for the underdoped case ($n=\filla$) and middle row for the overdoped case ($n=\fillb$). In both cases the gap function for the leading eigenvalue is a $d_{x^2-y^2}$ wave which is followed by another $d$-wave state with $d_{xy}$ symmetry and an $s$-wave state. Note that the gap function has been normalized such that $\mathcal N=1$ in Eq.~(\ref{eq_normal}).
Bottom row: Bottom row: Cuts at $k_z=\pi$ and $k_z=2\pi$ for the overdoped case to show horizontal accidental nodes in the leading $d_{x^2-y^2}$-wave gap (g, h), vertical nodes (i, j) and loop-like accidental nodes (k, l) on the \textit{M}-centered pockets.
 \label{fig_gapfunctions}}
\end{figure*}

 While the  model calculations\cite{Khodas12,Hu13B,*Hu13a} beautifully elucidated some of the factors controlling the superconducting ground state,  it is still of great interest to explore where in the  model phase diagram of Ref. \onlinecite{Khodas12} real materials such as K$_x$Fe$_{2-y}$Se$_2$ are located, and to compare with experiments, to see if this approach can be justified.
We first explore the magnetic susceptibility within our model of the band structure, since it reflects the physics of the spin-fluctuation pairing interaction.
In order to get a large RPA enhanced susceptibility, we choose the spin-rotational invariant interaction parameters $U=0.88\;\text{eV}$, $J=U/4$, a choice that leads to reasonable eigenvalues for pairing strength and still allows one to carry out the calculations with both fillings while avoiding the magnetic instability.
We find that the total magnetic susceptibilities have little $q_z$ dependence; results for $\chi_0(q_x,q_x,0)$ and the RPA spin susceptibilities $\chi_{\text{RPA}}(q_x,q_y,0)$ are shown in Fig.~\ref{fig_susc}.
The flat plateau in $\chi_0$ around the $X$ point is enhanced by
interactions leading to a peak structure in the RPA susceptibility at an
incommensurate wave vector $\mathbf q\approx\pi(1.65,0.35,q_z)$ for
$n=\filla$ and  $\mathbf q\approx\pi(1.6,0.4,q_z)$ for $n=\fillb$.
These are both close to the wave vector determined in early 2D calculations\cite{Maier11}, as well as to the
 wave vector of the neutron resonance peak in RbFe$_2$Se$_2$ determined by Inosov \textit{et al.}\cite{Inosov11,Inosov12}.   In the underdoped case the system is closer to the spin-density wave instability than in the overdoped case. The latter  also has a slightly smaller bare susceptibility $\chi_0(\mathbf q=0)$, reflecting a lower density of states at the Fermi level.

\begin{figure*}[tb]
 \includegraphics{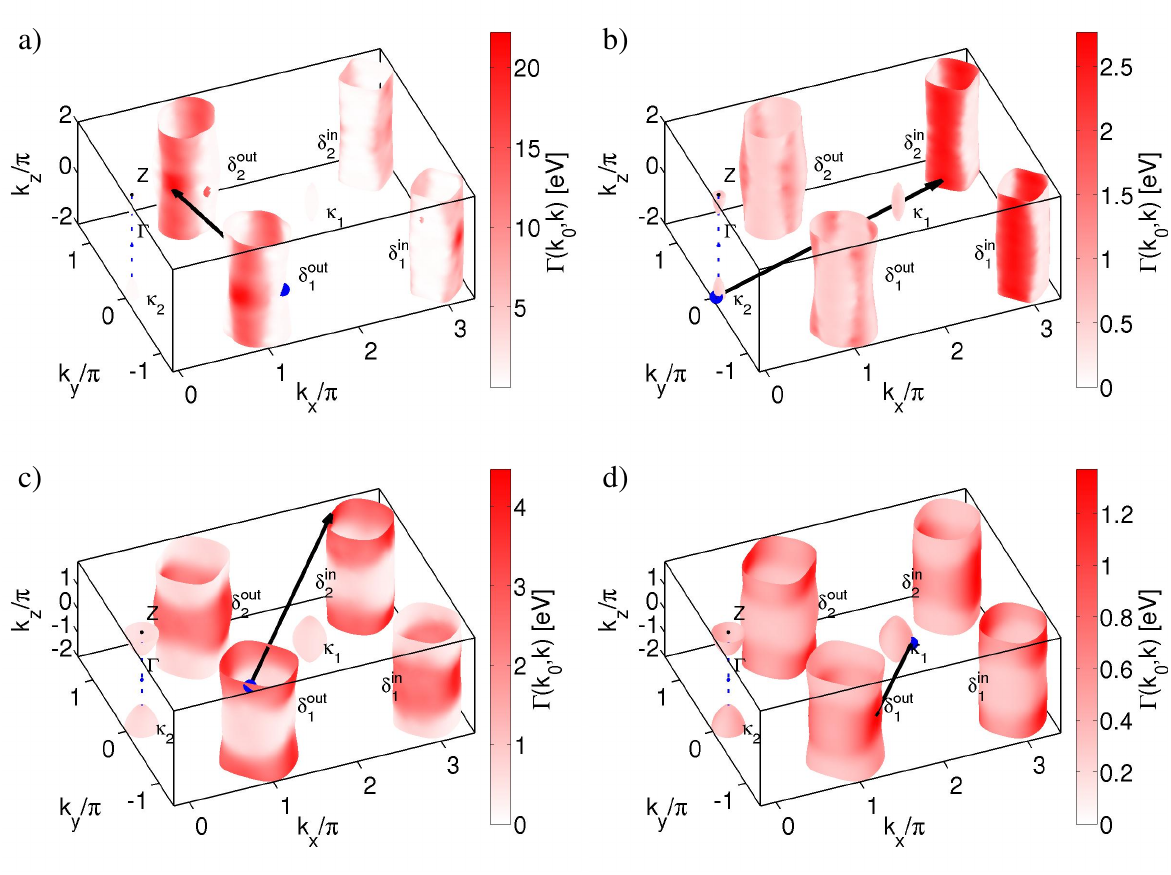}
\caption{(Color online) Calculated pair scattering vertex $\Gamma_{ij}(\mathbf k_0, \mathbf k)$ where $\mathbf k_0$ is fixed at one point of the Fermi surface [thick blue (dark gray) point] and the surface is colored as a function of $\bf k$  according to the value of the scattering vertex calculated using rotational invariant interaction parameters with $U=0.88\;\text{eV}$ and $J=U/4$. (a), (b) Two different $\bf k_0$ on a $M$-centered $\delta$ pocket and a $Z$-centered $\kappa$ pocket, respectively,  in the underdoped case; (c), (d) same as (a), (b), but in the overdoped case.    The blue points $\mathbf k_0$ are chosen to correspond to the (a), (c) maximum and (b), (d) minimum of the scattering vertex $\Gamma_{\k_0,\k'}$ over all $\k'$ in the Brillouin zone.
 The black arrows then indicate in each case  the largest scattering processes starting from $\k_0$.  Note that the scales of the four panels are very different and the inner pockets are artificially shifted in the $k_x$ direction by $2\pi$.
 \label{fig4}}
\end{figure*}
We now calculate the  spin-singlet pair vertex by symmetrizing the opposite-spin  pairing vertex $\Gamma_{ij}(\mathbf k, \mathbf k')\rightarrow
1/2[\Gamma_{ij}(\mathbf k, \mathbf k')+\Gamma_{ij}(\mathbf k, -\mathbf k')]$ using Eq.~(\ref{eq:Gam_ij}).  The leading eigenfunctions obtained by solving Eq.~(\ref{eqn:lingap1})
are then presented for the two doping cases in Fig.~\ref{fig_gapfunctions}.  One notices first that the order of the leading pairing channels is the same  in both cases.  A $d_{x^2-y^2}$ state is the clear dominant pairing instability; while the 3D gap functions are complicated, the gaps on the $\delta_1$ pockets are seen to transform into minus those on the $\delta_2$ pockets under $\pi/2$ rotations of the system about the $\Gamma$-$Z$ axis.  The same behavior is observed for the second leading pair state in both cases, which is however easily identifiable as a $d_{xy}$ state by the symmetry-enforced vertical nodal lines along the $(0,0,k_z)$-$(\pi,\pi,k_z)$ directions.  The third leading pair state is of $s$-wave symmetry, as verified by inspection of the behavior of the $\delta$ pockets under $\pi/2$ rotations, and by confirming that the $Z$-centered $\kappa$ pocket has no nodes.    In addition, one sees that the sign of the gap on the inner
$\delta$ is opposite to that of the gap on the outer $\delta$ pocket, so that from a symmetry point of view this state is indeed of the bonding-antibonding $s_\pm$ type. 
However, we note that its structure is considerably more complicated than the simple 2D models, and that this gap displays strong anisotropy over the electron pockets.   In the underdoped case, deep vertical minima are observed, while in the overdoped case loop nodes are formed.  While Ref. \onlinecite{Khodas12_PRB} includes some discussion of the momentum dependence of the hybridization term on the orbital physics, here we can see by explicit comparison with  Fig. \ref{fig1} that the orbital weights vary strongly over the entire Fermi surface and strongly influence  the overall 3D pairing interaction in the realistic case. Finally, we note that for fixed interaction parameters,  the pairing strength for the underdoped case is larger ($\lambda_1=1.9$) than for the overdoped case ($\lambda_1=0.54$). This simply reflects the fact that the underdoped system ($n=6.12$) is closer to the magnetic instability for the chosen parameters.
These results are quite robust with respect to changing either the  interaction parameters or  tight-binding parameters, provided that the band structure near the Fermi surface is not too different from that observed in ARPES experiments\cite{Qian11,Chen11}.

We now analyze these results by examining the structure of the pair vertex in each doping case to identify the predominant pair scattering processes.
In Fig. \ref{fig4}, we fix the wave vector of the  initial pair $(\k_0,-\k_0)$ ($\k_0$ is indicated by a thick blue dot), and plot the  vertex  as a function of the wave vector $\k$ of the scattered pair $(\k,-\k)$ .  In both cases, the predominant scattering processes are located at wave vectors associated with the
peaks in the susceptibility (Fig. \ref{fig_susc}), produced by scattering between the $\delta$-electron pockets as expected.  However the three-dimensional character of the dominant processes varies from case to case, determined by the higher amplitude of scattering between states of like orbital character\cite{Maier09}, as can be seen by comparison with Fig. \ref{fig1}.    In the case of the underdoped system, the scattering is seen to be mostly
of intra-orbital type between the $d_{xy}$ states on the outer electron pockets for $q_z=0$.  For the overdoped system, the finite $q_z$ interorbital $d_{xy}$-$d_{yz}$ scattering processes between the outer $\delta$ pockets appear to dominate, with smaller but comparable processes between the same orbitals.   While the dominant processes in both cases are indeed between $M$-centered electron pockets, they are somewhat different from those deduced from the simpler 2D models\cite{Fang11,Maier11}, highlighting the importance of both orbital weight effects and 3D scattering effects.  By the
latter we mean processes with scattering vector $\k_0-\k$ with a nonzero $z$ component corresponding to significant $\Gamma_{ij}(\k_0,\k)$; these are seen clearly in Fig. \ref{fig4}.
The influence of the $Z$-centered $\kappa$ electron pocket    is rather small for  both cases (giving small gap amplitudes) which is not surprising since the magnitude of the pairing vertex $\Gamma_{ij}(\mathbf k,\mathbf k')$ is much smaller and the $\kappa$ pockets themselves have a small density of states compared to the $\delta$ pockets.  Nevertheless it is worth noting that the subleading bonding-antibonding $s_\pm$-wave
instability takes advantage of pair scattering from this pocket in the overdoped case to create a  large gap value on the $\kappa$ sheets, as seen in Fig.~\ref{fig_gapfunctions}.

\begin{figure}
 \includegraphics[width=\linewidth]{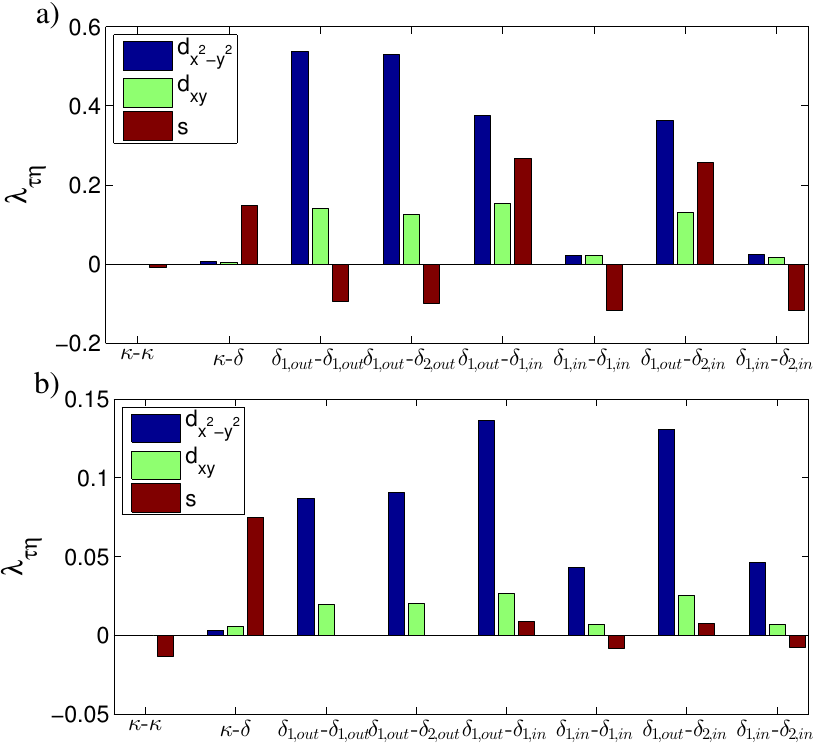}
 \caption{(Color online) Comparison of the partial contributions to the eigenvalue $\lambda_{ij}$ from scattering between the different pockets for the three leading gapfunctions at (a) $n=\filla$ and (b) $n=\fillb$.
$\lambda_{\kappa,\kappa}$, $\lambda_{\kappa,\delta}$
and then
$\lambda_{\delta_1,\mathrm{out};\delta_1,\mathrm{out}}$,  $\lambda_{\delta_1,\mathrm{out};\delta_2,\mathrm{out}}$,$\lambda_{\delta_1,\mathrm{out};\delta_1,\mathrm{in}}$,$\lambda_{\delta_1,\mathrm{in};\delta_1,\mathrm{in}}$,
$\lambda_{\delta_1,\mathrm{out};\delta_2,\mathrm{in}}$}
 \label{fig_lij_comp}
\end{figure}

In Fig.~\ref{fig_lij_comp}, we now highlight the contributions $\lambda_{\tau\eta}$  to the pairing of the various intraband and interband processes. Note that the contributions from scattering between the $\kappa$ pockets and between $\kappa$ pockets and $\delta$ pockets have been summed up ($\lambda_{\kappa\kappa}=2\lambda_{\kappa_1\kappa_1}+2\lambda_{\kappa_1\kappa_2}$ and $\lambda_{\kappa\delta}=[(\lambda_{\kappa_1\delta_1,\mathrm{in}}+\lambda_{\kappa_1\delta_1,\mathrm{out}}+ \delta_1\leftrightarrow \delta_2)+\kappa_1\leftrightarrow \kappa_2])$ since the overall value is small while for all other contributions we break the sum Eq.~(\ref{eq:lambda_explicit}) down to contributions between all pockets. The small differences between contributions of first or second $\delta$ pocket, (for example $\lambda_{\delta_1,\mathrm{in};\delta_1,\mathrm{out}}$ and $\lambda_{\delta_1,\mathrm{in};\delta_2,\mathrm{out}}$) are the result of the fact that the susceptibility and therefore also the pairing vertex is periodic in the  1-Fe Brillouin zone.
We note that, as expected, the $\delta$-$\delta$ pair scattering processes dominate, but some interesting subtleties are also seen.  As the system is
underdoped, the processes involving the outer pockets, both intraband and interband are the only ones which make a significant contribution, whereas for the overdoped cases all processes within and between both $\delta$ pockets contribute roughly equally.
Generally the $d$-wave state has larger $\lambda$ because it not only gains pairing strength from processes connecting inner to outer pockets, but also from processes connecting inner to inner and outer to outer pockets, for which the $s$-wave state has negative contributions.

\section{Effect of band hybridization}
\label{sec:hybrid}
\begin{figure*}
 \includegraphics[width=\linewidth]{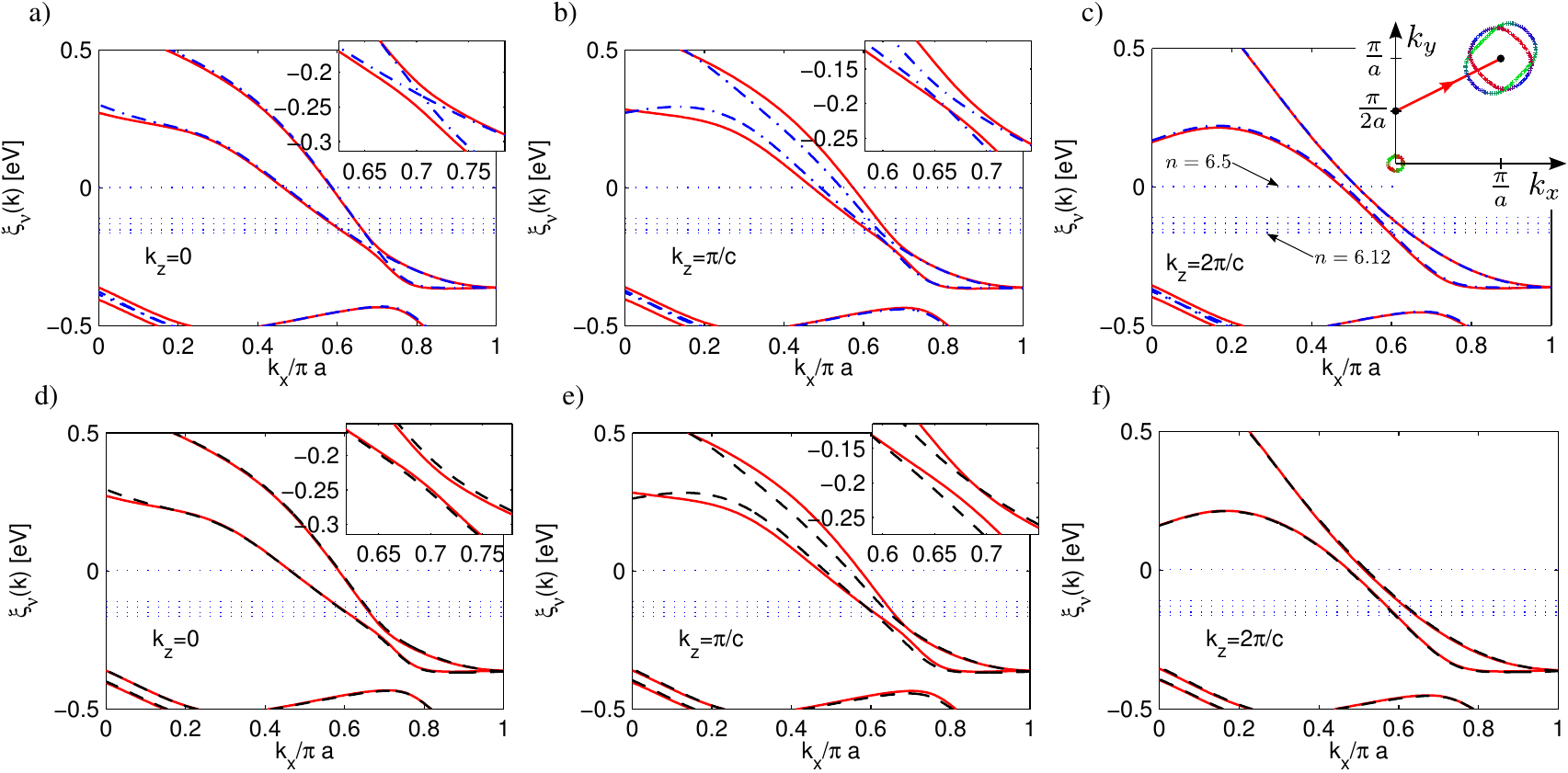}
 \caption{(Color online) Comparison of bands within different approximations shown along low symmetry cuts: Full relativistic DFT including spin-orbit coupling (solid, red) compared to the tight-binding bands without spin-orbit coupling at (a) $k_z=0$, (b) $k_z=\pi/c$ and (c) $k_z=2\pi/c$; full relativistic DFT compared to the tight-binding bands with spin-orbit coupling using $\lambda^{3d}_{\text{Fe}}=0.05\,\text{eV}$ at (d) $k_z=0$ and (e) $k_z=\pi/c$  and (f) $k_z=2\pi/c$. The insets show details close to the regions where the bands hybridize, and the dotted (blue) lines are the chemical potentials at fillings $n=6.12$, $6.15$, $6.20$, $6.25$\text{ and }$6.5$ (parent compound).}
 \label{fig_SOCbands}
\end{figure*}

Within our approach presented until now, we have adopted the results of a tight-binding Wannier downfolding of {\it nonrelativistic} DFT bands for KFe$_2$Se$_2$.
For systems in the  $P4/nmm$ symmetry class, high symmetry paths on the boundary of the tetragonal 2-Fe Brillouin zone are degenerate\cite{Eschrig09}.
However, for the 122 crystal structure (space group $I4/mmm$), these bands must hybridize and such degeneracies are lifted.  As discussed in
the Appendix,  these band splittings are still extremely small in this particular 122 system, of order $2\,\text{meV}$.  This is illustrated in Fig. \ref{fig_SOCbands}(a),
where this hybridization is essentially invisible.  Thus when one examines the inner and outer electron $\delta$-pockets around
these near-crossing points the change in orbital weight is particularly abrupt, as seen in Fig. \ref{fig1}.

\begin{figure}
 \includegraphics[width=\linewidth]{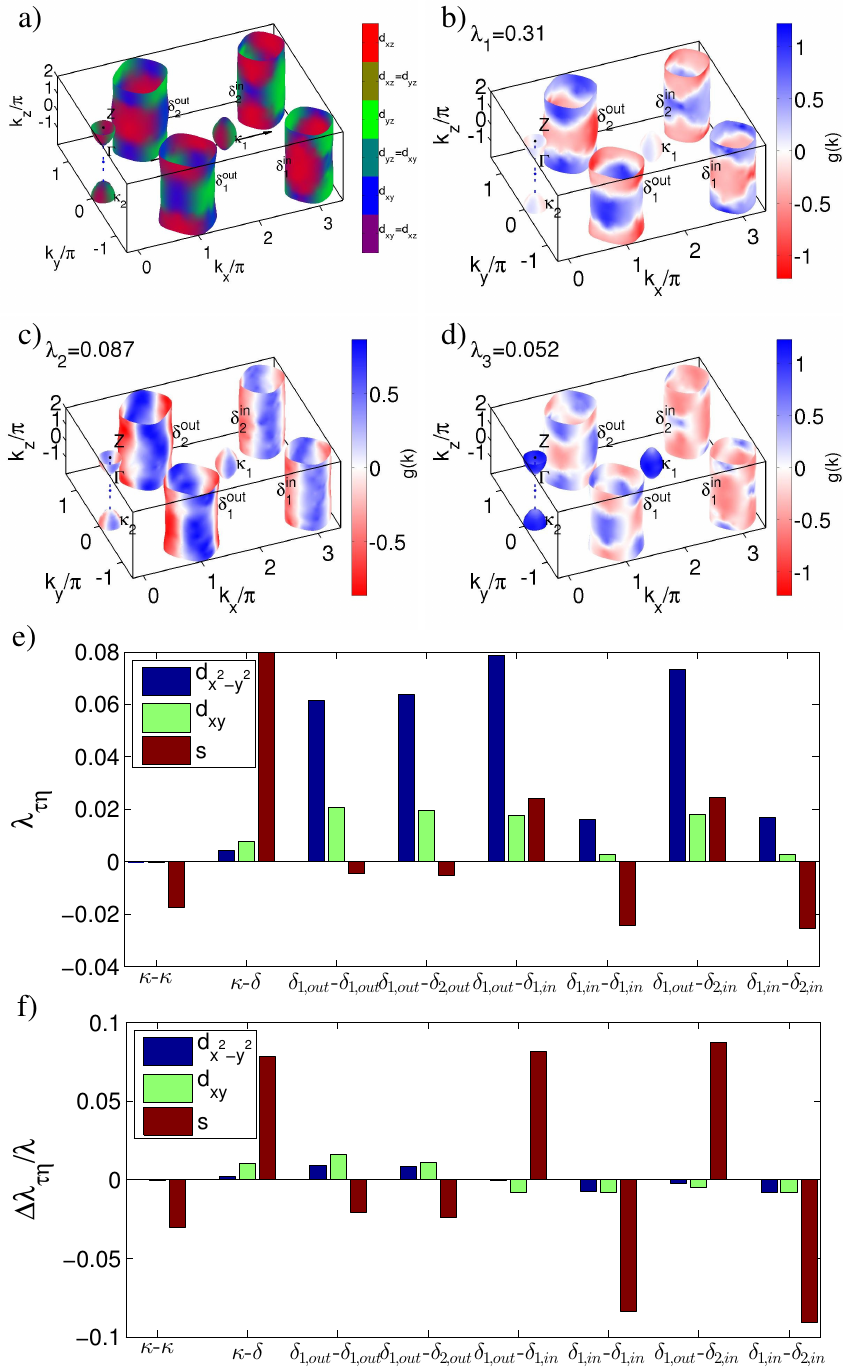}
 \caption{(Color online) Fermi surface (a) and leading pairing states including spin-orbit coupling for filling $n_2=\fillb$ with $U=0.88\,\text{eV}$ calculated with the spin-orbit coupling constant $\lambda^{3d}_{\text{Fe}}=0.05\,\text{eV}$ (b)-(d) together with the partial contributions to the pairing eigenvalue (e) and the relative changes of the partial contributions on switching on the spin-orbit coupling (f).}
 \label{fig_pairing_SO}
\end{figure}

Within the context of the fluctuation-exchange approximation and the treatment of the electronic structure described in Sec.~\ref{sec:Model}, we always find that the leading pair instability is in   the $d_{x^2-y^2}$ channel.
If one examines the predictions of the model approach of Khodas and Chubukov, one is led directly to the conclusion that these
results, for a wide range of dopings and interaction parameters, suggest that the system is in the small-$\kappa$, i.e. extremely weak hybridization part of the phase diagram. As noted by these authors, in the limit of large hybridization a bonding-antibonding $s_\pm$-wave state is favored.  One may ask, therefore, if the hybridization has been treated sufficiently accurately here, or in other words, can small changes in the accuracy of these bands  lead to dramatically different results, in particular the stability of the bonding-antibonding $s_\pm$-wave state?    To answer this question, we consider the effect of spin-orbit coupling which is neglected in the nonrelativistic treatment previously described, and was known to enhance the splitting of the electron bands.
To perform the
calculation for the pairing in a completely consistent relativistic approximation requires not only the calculation of the one-electron states in the presence of the spin orbit coupling,   but also the recalculation of the interaction matrices in the pseudospin basis in which the one-electron Hamiltonian  is diagonal.  It is therefore beyond the scope of
this paper, and we instead present an approximation based on the correct relativistic calculation of the band structure, but in which the eigenstates are projected on the major spin component, e.g. only the spin-up component after a renormalization is used.
The main purpose of this approximation is
simply to create a larger splitting of the $\delta$-electron pockets to see if a larger hybridization might change the order of the pairing symmetry states.
In order to take spin-orbit coupling into account we add a phenomenological spin-orbit coupling term
\begin{equation}
 H_{\text{SO}}=\lambda^{3d}_{\text{Fe}}\sum_i \sum_{\alpha=x,y,z} L_i^\alpha S_i^\alpha\,,\label{eq_H_so}
\end{equation}
where the spin-orbit coupling constant $\lambda^{3d}_{\text{Fe}}$  for Fe $3d$ orbitals can be obtained from the corresponding wavefunctions and the crystal potential\cite{Friedel64}. The matrices of the components of the angular momentum operator are now evaluated in the real basis and lead to a total { $20\times 20$} Hamiltonian in orbital and spin space.
The eigenenergies are then exactly the ones one would use in a full calculation due to the Kramers doublets; however the set of eigenstates { projected onto the majority spin subspace}  does not {exactly} describe the real quantum state.
The band energies obtained from Eq.~(\ref{eq_H_so}) using a single spin-orbit coupling constant $\lambda^{3d}_{\text{Fe}}=0.05\;\text{eV}$ match the DFT bands with spin-orbit coupling quite well, as  can be seen in Fig.~\ref{fig_SOCbands}, where details of bands from a relativistic (self-consistent) DFT calculation are compared to the results using the nonrelativistic band structure with the spin-orbit coupling term in Eq.~(\ref{eq_H_so}). Although there are deviations between the self-consistent DFT calculation (solid lines) and the approximation (dashed line), the splitting of the bands is quite similar.
In Fig.~\ref{fig_pairing_SO}(a) we present the Fermi surface derived from the model including spin-orbit coupling which shows stronger hybridization and mixing of orbital weights compared to the one in Fig.~\ref{fig1}(b).

The overall properties of the pairing states are mostly unaffected, but details do change.  One sees, for example, that the eigenvalues $\lambda_i$ are generally suppressed for fixed interaction parameters due to the Fermi surface changes. In addition, the partial contributions $\lambda_{\tau\eta}$ show similar features as seen in Fig.~\ref{fig_lij_comp}(b) where the scattering on the $\delta$ pockets again plays the dominant role. However, as seen in Fig.~\ref{fig_pairing_SO}(f), the partial contributions of the leading $d$-wave state are only altered by a few percent while the $\lambda_{\tau\eta}$ of the $s$-wave state have changed significantly.  As seen in the figure, the changes due to hybridization stabilize
the bonding-antibonding $s$-wave state via strengthening of interpocket processes\cite{Khodas12}, which also make the state more isotropic on
  the $\delta$ pockets.  The relative enhancement of the state by hybridization is, however, not
   as large as might have been expected,  due to the repulsive intrapocket processes, which cost energy as the nodes are suppressed    [Fig.~\ref{fig_pairing_SO}(f)].

 Our results above indicate strongly that the real K$_x$Fe$_{2-y}$Se$_2$ is  in the limit of small hybridization of the
phase diagram proposed by Khodas and Chubukov\cite{Khodas12}.  To verify this, we can estimate for our 3D system both the hybridization and
ellipticity entering the effective tuning parameter $\kappa_{\mathrm{eff}}$, yielding
\begin{equation}\label{eq:kappa}
  \kappa_{\mathrm{eff}} \simeq \frac{\delta\xi}{ \langle \frac{\Delta\xi_\k}{ v_\mathrm{F}(\k)k_\mathrm{F}(\k)} \rangle},
\end{equation}
where $\delta\xi$ is the maximum splitting of low symmetry degeneracies by hybridization, of order 10 meV, and $\Delta \xi $ is
the difference of the two electron  band energies crossing the Fermi surface near $M$, a measure of the ellipticity.  For
our Fermi surfaces, this yields values of $\kappa$ of order $0.1$-$0.3$, significantly below the crossover of order $\kappa\approx 1$ identified
 as the instability of the $d$-wave state, consistent with our results for the pairing state.

\section{Conclusions}
\label{sec:conclusions}
The K$_x$Fe$_{2-y}$Se$_2$ system is one of the most interesting of the Fe-based superconductors under current investigation, in part because
the lack of a $\Gamma$-centered hole pocket makes the usual $s_\pm$ gap structure less likely.   We have explored the possible pairing symmetries within a
full-scale 3D microscopic spin fluctuation-exchange calculation using a ten Fe-orbital tight-binding band structure which respects the $I4/mmm$ space group symmetry of the crystal, and found a variety of possible states.   For two different fillings consistent with the ARPES-determined absence of the hole pocket,
we  find that the $d_{x^2-y^2}$ symmetry is dominant, with a $d_{xy}$ state as the next leading instability.  These states are the 3D analogs of 2D nodeless $d$-wave states discussed in earlier calculations\cite{Maier11,Wang11,Kontani11,Maier12}, but as required by crystal symmetry\cite{Mazin11} display ``quasi-nodes" on the large electron pockets which may be vertical, horizontal, or looplike depending on filling.  The DFT band structure used to generate our tight-binding band structure in this first part of our analysis shows that this particular system exhibits extremely small hybridization of the electron pockets.   Following the proposal of Refs. \onlinecite{Mazin11} and \onlinecite{Khodas12}, where it was demonstrated that the bonding-antibonding $s_\pm$ state (which changes sign between the inner and outer electron pocket) is stabilized through hybridization, we attempted to increase the splitting of the bands by including spin-orbit coupling in our calculation in an approximation which agrees well with full relativistic DFT calculations.  We found that, while the stability of the $s$-wave state was indeed enhanced somewhat relative to the $d$-wave states,  the effect was too small to make  the third-place $s$-wave state dominant over the $d$ wave,  and we believe that the real system is in the small effective hybridization ($\kappa_{\mathrm{eff}}$)  limit where the $d$ wave is stable\cite{Khodas12}.  The width of the quasinodes of the $d$-wave gap function on the large electron pockets indeed appears to scale with the strength of the hybridization, as anticipated in
Ref.~\onlinecite{Maier11}.

 As further discussed in Ref. \onlinecite{Maier11}, since the $d$-wave gap functions found here vary so rapidly near the quasi-nodes on the large electron pockets, it is possible that the small quasiparticle phase space they imply would not be detectable in the low-resolution thermodynamic experiments which have concluded that this system is nodeless.  In this and other early analyses, the
 additional electronlike $Z$-centered Fermi pocket observed in the ARPES analysis of this system\cite{Xu12} was neglected; indeed, it is found here to have very little influence on the eigenvalues of the pairing states in our calculations.  Nevertheless, the gap function induced on these pockets is subject to the same requirements of pairing symmetry as the gap on the
 rest of the Fermi surface.  In particular, $d$-wave symmetry implies that nodes are required on these small pockets, and the report of an isotropic gap on this pocket by Xu \textit{et al.}\cite{Xu12} is indeed inconsistent with our finding that the $d$ wave is always dominant. It is possible that the gap anisotropy on this very small Fermi surface feature is harder to resolve than anticipated by the authors of Ref. \onlinecite{Xu12}, or that our calculations are missing ingredients (such as stronger correlations) which are important in the
chalcogenide systems.  It is also worth mentioning in closing that our starting point is based on a band structure associated with the doped KFe$_{2}$Se$_2$ compound, which
may be different from the actual composition and structure of the superconducting material\cite{HKM11,dagotto_review,hhwen_review}.

 \begin{acknowledgments}
The authors gratefully acknowledge useful discussions with T. Berlijn, A. Chubukov, I. Eremin, H. Jeschke, M. Khodas,  M. Korshunov, G. Panchapakesan, M.~Taillefumier, M. Tomi\'c and R. Valent\'i.
P.J.H., Y.W., and A.K. were supported by DOE DE-FG02-05ER46236.  A portion of this research was conducted at the Center for Nanophase Materials Sciences, which is sponsored at Oak Ridge National Laboratory by the Scientific User Facilities Division, Office of Basic Energy Sciences, U.S. Department of Energy.
\end{acknowledgments}

\section*{APPENDIX}
\subsection{DFT derived bandstructure of the parent compound KFe$_2$Se$_2$}
All DFT calculations performed in this work are based on the WIEN2k\cite{Blaha2001} package using the generalized gradient approximation [Perdew-Burke-Ernzerhof (PBE) functional]\cite{PBE} where the structure parameters for KFe$_2$Se$_2$ (space group $I4/mmm$ at $T = 297\,\text{K}$) provided in Ref.~\onlinecite{Guo10} were used.  The lattice constants are $a = 3.9136\,\text{\AA}$,
$c = 14.0367\,\text{\AA}$ and the internal coordinates for the Se atoms were fixed at $z_{\text{Se}} = 0.3539$.
\subsubsection{Band structure and Wannier projection without spin-orbit coupling}
 \begin{figure}[tb]
 \includegraphics{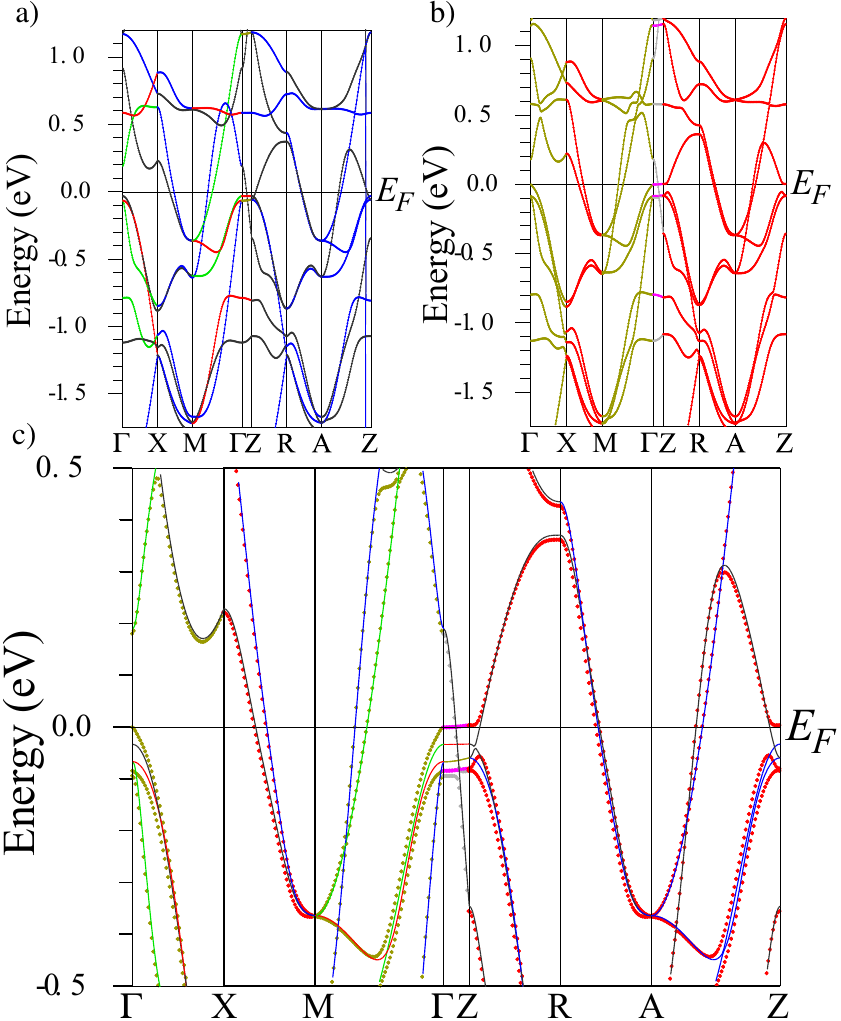}
\caption{(Color online) DFT band structure (a) without spin-orbit coupling and (b) with spin-orbit coupling and (c) a direct comparison of the two results (dots: with spin-orbit coupling). The color codes the symmetries according to the irreducible representation of the point group.
 \label{Fig_bands_dft}}
\end{figure}

Good convergence is obtained for a $k$ mesh with 3000 $k$ points ($14^3$) that was checked up to $27^3$ $k$ points together with $R_{\mathrm{MT}}\times K_{\mathrm{MAX}}$ of $8.0$, the band structure can be seen in Fig.~\ref{Fig_bands_dft}(a).
 For the Wannier projection onto a ten orbital tight-binding model the $2\times5$ $d$ orbitals were selected as an initial guess and an outer energy window of  $-2.3 \ldots 3.6\,\text{eV}$ was selected in conjunction with an inner window of $-0.5 \ldots 0.3 \,\text{eV}$ to obtain reasonable agreement in the low-energy regime. The subsequent projection on atomic orbitals and the calculation of maximally localized Wannier functions yield the initial tight-binding hoppings for the single particle Hamiltonian Eq.~(\ref{eq_H0}). For completeness we show results of the band structure and the Fermi surface in Figs.~\ref{Fig_bands_dft}(a) and \ref{Fig_FS_dft}(a). These compare well to previous investigations\cite{Shein11}.
 \subsubsection{Band structure with spin-orbit interaction}
 \begin{figure}[tb]
\includegraphics{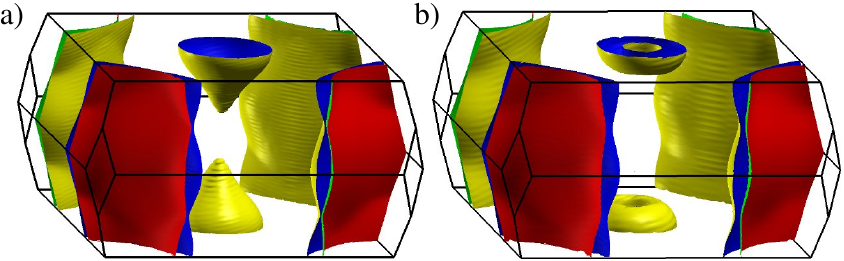}
\caption{(Color online) Fermi surface generated using the DFT band structure (a) without spin-orbit coupling and (b) with spin-orbit coupling.
 \label{Fig_FS_dft}}
\end{figure}
Although the elements in the compound under investigation are of relatively small atomic number $Z$, relativistic effects can modify the Fermi surface and thus influence the pairing state. Especially the hybridization of the $M$-centered electronlike pockets gets enhanced which may stabilize the $s_\pm$-wave state. To check this qualitatively and quantitatively, we start from the results of the DFT calculation and add the spin-orbit coupling on the Fe atoms. The convergence is checked for up to 7500 $k$ points together with  $R_{\mathrm{MT}}\times K_{\mathrm{MAX}}$ up to $9.0$.\cite{Cottenier,Note2}
A Wannier projection of the DFT bandstructure (onto a 20-orbital model) together with a pairing calculation in the pseudospin space is beyond the scope of this paper. However, we use the wavefunction at convergence to estimate the spin-orbit coupling constants $\lambda^{3d}_\text{Fe}\approx 0.06\,\text{eV}$ ($\lambda^{4p}_\text{Se}\approx 0.5\,\text{eV}$), where the first enters into the spin-orbit Hamiltonian Eq.~(\ref{eq_H_so}) to give an accurate description of the band splittings close to the Fermi level. For completeness, we show the band structure and the Fermi surface of the parent compound in the presence of spin-orbit coupling in Figs.~\ref{Fig_bands_dft}(b) and \ref{Fig_FS_dft}(b).
 \subsection{Phenomenological spin-orbit coupling}
 \begin{figure}
 \includegraphics[width=\linewidth]{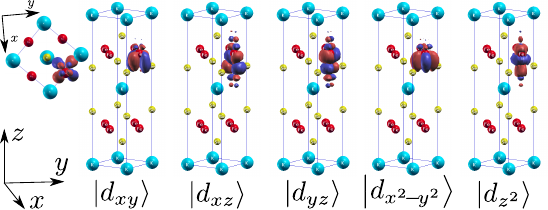}
 \caption{(Color online) Wannier functions centered at the first Fe atom visualized in the conventional elementary cell. The major atomic character that is used to calculate the contribution to the spin-orbit coupling is indicated by the states below the individual pictures.}
 \label{Fig_wan}
\end{figure}
 Considering spin-orbit coupling on all states on all atoms in a crystal one can write the additional term in the Hamiltonian as
 \begin{equation}
  H_{\text{SO}}^{\text{full}}=\sum_s \sum_n \sum_{l=1}^{n} \lambda_s^{nl}  \left. \mathbf L\right|_l \cdot \mathbf S\,,\label{eq_so_full}
 \end{equation}
 where the sums run over all atoms $s$, principal quantum numbers $n$ and angular quantum numbers $l$; the $l=0$ term has already been dropped since it is zero anyhow.
 $\left. \mathbf L\right|_l$ is the angular momentum operator for fixed $l$ and $\lambda_s^{nl}$ is the corresponding spin-orbit coupling constant. We are interested in the effects on the electronic structure close to the Fermi level and thus can simplify Eq.~(\ref{eq_so_full}) to
 \begin{equation}
  H_{\text{SO}}^{\text{red}}=\lambda^{3d}_\text{Fe}\left. \mathbf L\right|_2 \cdot \mathbf S\,,
 \end{equation}
 where only the spin-orbit constant of the Fe atoms for $l=2$ enters since the states at the Fermi level are mainly of Fe \mbox{$d$-character}. Looking at the Wannier functions which are shown for one Fe atom in Fig.~\ref{Fig_wan} one can approximate these by the real atom basis states $\{\left|z^2\right>,\left|xz\right>,\left|yz\right>,\left|x^2-y^2\right>,\left|xy\right>\}$ in order to evaluate the matrix elements of the angular momentum operator, giving
  \begin{subequations}
   \begin{equation}
  L^z=\left(\begin{array}{ccccc}
             0&0&0&0&0\\
             0&0&-i&0&0\\
             0&i&0&0&0\\
             0&0&0&0&-2i\\
             0&0&0&2i&0
            \end{array}\right),
 \end{equation}
  \begin{equation}
  L^x=\left(\begin{array}{ccccc}
             0&0&\sqrt 3 i&0&0\\
             0&0&0&0&i\\
             -\sqrt 3 i&0&0&-i&0\\
             0&0&i&0&0\\
             0&-i&0&0&0
            \end{array}\right),
 \end{equation}
 \begin{equation}
   L^y=\left(\begin{array}{ccccc}
             0&\sqrt3i&0&0&0\\
             -\sqrt 3i&0&0&i&0\\
             0&0&0&0&i\\
             0&-i&0&0&0\\
             0&0&-i&0&0
            \end{array}\right) \,.\label{eq_Limatrices}
 \end{equation}
 \end{subequations}
These enter into the spin-orbit coupling term Eq.~(\ref{eq_H_so}), where the index for $l=2$ has been dropped.

%


\begin{thebibliography}{10}%
\makeatletter
\providecommand \@ifxundefined [1]{%
 \ifx #1\undefined \expandafter \@firstoftwo
 \else \expandafter \@secondoftwo
\fi
}%
\providecommand \@ifnum [1]{%
 \ifnum #1\expandafter \@firstoftwo
 \else \expandafter \@secondoftwo
\fi
}%
\providecommand \enquote [1]{``#1''}%
\providecommand \bibnamefont  [1]{#1}%
\providecommand \bibfnamefont [1]{#1}%
\providecommand \citenamefont [1]{#1}%
\providecommand\href[0]{\@sanitize\@href}%
\providecommand\@href[1]{\endgroup\@@startlink{#1}\endgroup\@@href}%
\providecommand\@@href[1]{#1\@@endlink}%
\providecommand \@sanitize [0]{\begingroup\catcode`\&12\catcode`\#12\relax}%
\@ifxundefined \pdfoutput {\@firstoftwo}{%
 \@ifnum{\z@=\pdfoutput}{\@firstoftwo}{\@secondoftwo}%
}{%
 \providecommand\@@startlink[1]{\leavevmode}%
 \providecommand\@@endlink[0]{}%
}{%
 \providecommand\@@startlink[1]{%
  \leavevmode
  \pdfstartlink
   attr{/Border[0 0 1 ]/H/I/C[0 1 1]}%
   user{/Subtype/Link/A<</Type/Action/S/URI/URI(#1)>>}%
  \relax
 }%
 \providecommand\@@endlink[0]{\pdfendlink}%
}%
\providecommand \url  [0]{\begingroup\@sanitize \@url }%
\providecommand \@url [1]{\endgroup\@href {#1}{\urlprefix}}%
\providecommand \urlprefix [0]{URL }%
\providecommand \Eprint[0]{\href }%
\@ifxundefined \urlstyle {%
  \providecommand \doi [1]{doi:\discretionary{}{}{}#1}%
}{%
  \providecommand \doi [0]{doi:\discretionary{}{}{}\begingroup
  \urlstyle{rm}\Url }%
}%
\providecommand \doibase [0]{http://dx.doi.org/}%
\providecommand \Doi[1]{\href{\doibase#1}}%
\providecommand \bibAnnote [3]{%
  \BibitemShut{#1}%
  \begin{quotation}\noindent
    \textsc{Key:}\ #2\\\textsc{Annotation:}\ #3%
  \end{quotation}%
}%
\providecommand \bibAnnoteFile [2]{%
  \IfFileExists{#2}{\bibAnnote {#1} {#2} {\input{#2}}}{}%
}%
\providecommand \typeout [0]{\immediate \write \m@ne }%
\providecommand \selectlanguage [0]{\@gobble}%
\providecommand \bibinfo [0]{\@secondoftwo}%
\providecommand \bibfield [0]{\@secondoftwo}%
\providecommand \translation [1]{[#1]}%
\providecommand \BibitemOpen[0]{}%
\providecommand \bibitemStop [0]{}%
\providecommand \bibitemNoStop [0]{.\EOS\space}%
\providecommand \EOS [0]{\spacefactor3000\relax}%
\providecommand \BibitemShut [1]{\csname bibitem#1\endcsname}%
\bibitem{stewart_review}%
  \BibitemOpen
  \bibfield{author}{%
  \bibinfo {author} {\bibfnamefont{G.}~\bibnamefont{Stewart}},\ }%
  \bibfield{journal}{%
  \bibinfo {journal} {Rev. Mod. Phys.}\ }%
  \textbf{\bibinfo {volume} {83}},\ \bibinfo {pages} {1589} (\bibinfo {year}
  {2011})%
  \bibAnnoteFile{NoStop}{stewart_review}%
\bibitem{HKM11}%
  \BibitemOpen
  \bibfield{author}{%
  \bibinfo {author} {\bibfnamefont{P.~J.}\ \bibnamefont{Hirschfeld}}, \bibinfo
  {author} {\bibfnamefont{M.~M.}\ \bibnamefont{Korshunov}},\ and\ \bibinfo
  {author} {\bibfnamefont{I.~I.}\ \bibnamefont{Mazin}},\ }%
  \bibfield{journal}{%
  \bibinfo {journal} {Rep. Prog. Phys.}\ }%
  \textbf{\bibinfo {volume} {74}},\ \bibinfo {pages} {124508} (\bibinfo {year}
  {2011})%
  \bibAnnoteFile{NoStop}{HKM11}%
\bibitem{dagotto_review}%
  \BibitemOpen
  \bibfield{author}{%
  \bibinfo {author} {\bibfnamefont{E.}~\bibnamefont{{Dagotto}}},\ }%
  \bibfield{journal}{%
  \bibinfo {journal} {Rev. Mod. Phys.}\ }%
  \textbf{\bibinfo {volume} {85}},\ \bibinfo {pages} {849} (\bibinfo {year}
  {2013})%
  \bibAnnoteFile{NoStop}{dagotto_review}%
\bibitem{hhwen_review}%
  \BibitemOpen
  \bibfield{author}{%
  \bibinfo {author} {\bibfnamefont{H.-H.}\ \bibnamefont{Wen}},\ }%
  \bibfield{journal}{%
  \bibinfo {journal} {Rep. Prog. Phys.}\ }%
  \textbf{\bibinfo {volume} {75}},\ \bibinfo {pages} {112501} (\bibinfo {year}
  {2012})%
  \bibAnnoteFile{NoStop}{hhwen_review}%
\bibitem{Qian11}%
  \BibitemOpen
  \bibfield{author}{%
  \bibinfo {author} {\bibfnamefont{T.}~\bibnamefont{Qian}}, \emph{et~al.},\ }%
  \bibfield{journal}{%
  \bibinfo {journal} {Phys. Rev. Lett.}\ }%
  \textbf{\bibinfo {volume} {106}},\ \bibinfo {pages} {187001} (\bibinfo {year}
  {2011})%
  \bibAnnoteFile{NoStop}{Qian11}%
\bibitem{Wang11}%
  \BibitemOpen
  \bibfield{author}{%
  \bibinfo {author} {\bibfnamefont{F.}~\bibnamefont{Wang}}, \emph{et~al.},\ }%
  \bibfield{journal}{%
  \bibinfo {journal} {Europhys. Lett.)}\ }%
  \textbf{\bibinfo {volume} {93}},\ \bibinfo {pages} {57003} (\bibinfo {year}
  {2011})%
  \bibAnnoteFile{NoStop}{Wang11}%
\bibitem{Maier11}%
  \BibitemOpen
  \bibfield{author}{%
  \bibinfo {author} {\bibfnamefont{T.~A.}\ \bibnamefont{Maier}}, \bibinfo
  {author} {\bibfnamefont{S.}~\bibnamefont{Graser}}, \bibinfo {author}
  {\bibfnamefont{P.~J.}\ \bibnamefont{Hirschfeld}},\ and\ \bibinfo {author}
  {\bibfnamefont{D.~J.}\ \bibnamefont{Scalapino}},\ }%
  \bibfield{journal}{%
  \bibinfo {journal} {Phys. Rev. B}\ }%
  \textbf{\bibinfo {volume} {83}},\ \bibinfo {pages} {100515} (\bibinfo {year}
  {2011})%
  \bibAnnoteFile{NoStop}{Maier11}%
\bibitem{Maiti}%
  \BibitemOpen
  \bibfield{author}{%
  \bibinfo {author} {\bibfnamefont{S.}~\bibnamefont{Maiti}}, \bibinfo {author}
  {\bibfnamefont{M.~M.}\ \bibnamefont{Korshunov}}, \bibinfo {author}
  {\bibfnamefont{T.~A.}\ \bibnamefont{Maier}}, \bibinfo {author}
  {\bibfnamefont{P.~J.}\ \bibnamefont{Hirschfeld}},\ and\ \bibinfo {author}
  {\bibfnamefont{A.~V.}\ \bibnamefont{Chubukov}},\ }%
  \bibfield{journal}{%
  \bibinfo {journal} {Phys. Rev. Lett.}\ }%
  \textbf{\bibinfo {volume} {107}},\ \bibinfo {pages} {147002} (\bibinfo {year}
  {2011})%
  \bibAnnoteFile{NoStop}{Maiti}%
\bibitem{Mazin11}%
  \BibitemOpen
  \bibfield{author}{%
  \bibinfo {author} {\bibfnamefont{I.~I.}\ \bibnamefont{Mazin}},\ }%
  \bibfield{journal}{%
  \bibinfo {journal} {Phys. Rev. B}\ }%
  \textbf{\bibinfo {volume} {84}},\ \bibinfo {pages} {024529} (\bibinfo {year}
  {2011})%
  \bibAnnoteFile{NoStop}{Mazin11}%
\bibitem{Khodas12}%
  \BibitemOpen
  \bibfield{author}{%
  \bibinfo {author} {\bibfnamefont{M.}~\bibnamefont{Khodas}}\ and\ \bibinfo
  {author} {\bibfnamefont{A.~V.}\ \bibnamefont{Chubukov}},\ }%
  \bibfield{journal}{%
  \bibinfo {journal} {Phys. Rev. Lett.}\ }%
  \textbf{\bibinfo {volume} {108}},\ \bibinfo {pages} {247003} (\bibinfo {year}
  {2012})%
  \bibAnnoteFile{NoStop}{Khodas12}%
\bibitem{Kontani11}%
  \BibitemOpen
  \bibfield{author}{%
  \bibinfo {author} {\bibfnamefont{T.}~\bibnamefont{Saito}}, \bibinfo {author}
  {\bibfnamefont{S.}~\bibnamefont{Onari}},\ and\ \bibinfo {author}
  {\bibfnamefont{H.}~\bibnamefont{Kontani}},\ }%
  \bibfield{journal}{%
  \bibinfo {journal} {Phys. Rev. B}\ }%
  \textbf{\bibinfo {volume} {83}},\ \bibinfo {pages} {140512} (\bibinfo {year}
  {2011})%
  \bibAnnoteFile{NoStop}{Kontani11}%
\bibitem{Fang11}%
  \BibitemOpen
  \bibfield{author}{%
  \bibinfo {author} {\bibfnamefont{C.}~\bibnamefont{Fang}}, \bibinfo {author}
  {\bibfnamefont{Y.-L.}\ \bibnamefont{Wu}}, \bibinfo {author}
  {\bibfnamefont{R.}~\bibnamefont{Thomale}}, \bibinfo {author}
  {\bibfnamefont{B.~A.}\ \bibnamefont{Bernevig}},\ and\ \bibinfo {author}
  {\bibfnamefont{J.}~\bibnamefont{Hu}},\ }%
  \bibfield{journal}{%
  \bibinfo {journal} {Phys. Rev. X}\ }%
  \textbf{\bibinfo {volume} {1}},\ \bibinfo {pages} {011009} (\bibinfo {year}
  {2011})%
  \bibAnnoteFile{NoStop}{Fang11}%
\bibitem{WBao_2011}%
  \BibitemOpen
  \bibfield{author}{%
  \bibinfo {author} {\bibfnamefont{B.}~\bibnamefont{Wei}}, \emph{et~al.},\ }%
  \bibfield{journal}{%
  \bibinfo {journal} {Chin. Phys. Lett.}\ }%
  \textbf{\bibinfo {volume} {28}},\ \bibinfo {eid} {86104} (\bibinfo {year}
  {2011})%
  \bibAnnoteFile{NoStop}{WBao_2011}%
\bibitem{Berlijn_2012}%
  \BibitemOpen
  \bibfield{author}{%
  \bibinfo {author} {\bibfnamefont{T.}~\bibnamefont{Berlijn}}, \bibinfo
  {author} {\bibfnamefont{P.~J.}\ \bibnamefont{Hirschfeld}},\ and\ \bibinfo
  {author} {\bibfnamefont{W.}~\bibnamefont{Ku}},\ }%
  \bibfield{journal}{%
  \bibinfo {journal} {Phys. Rev. Lett.}\ }%
  \textbf{\bibinfo {volume} {109}},\ \bibinfo {pages} {147003} (\bibinfo {year}
  {2012})%
  \bibAnnoteFile{NoStop}{Berlijn_2012}%
\bibitem{Blaha2001}%
  \BibitemOpen
  \bibfield{author}{%
  \bibinfo {author} {\bibfnamefont{P.}~\bibnamefont{Blaha}}, \bibinfo {author}
  {\bibfnamefont{K.}~\bibnamefont{Schwarz}}, \bibinfo {author}
  {\bibfnamefont{G.~K.~H.}\ \bibnamefont{Madsen}}, \bibinfo {author}
  {\bibfnamefont{D.}~\bibnamefont{Kvasnicka}},\ and\ \bibinfo {author}
  {\bibfnamefont{J.}~\bibnamefont{Luitz}},\ }%
  \bibfield{journal}{%
  \bibinfo {journal} {WIEN2k, \textit{An Augmented Plane Wave+Local Orbitals
  Program for Calculating Crystal Properties}}}%
   (\bibinfo {year} {Karlheinz Schwarz Techn. Universit\"at Wien, Austria
  2001})%
  \bibAnnoteFile{NoStop}{Blaha2001}%
\bibitem{Chen11}%
  \BibitemOpen
  \bibfield{author}{%
  \bibinfo {author} {\bibfnamefont{F.}~\bibnamefont{Chen}}, \emph{et~al.},\ }%
  \bibfield{journal}{%
  \bibinfo {journal} {Phys. Rev. X}\ }%
  \textbf{\bibinfo {volume} {1}},\ \bibinfo {pages} {021020} (\bibinfo {year}
  {2011})%
  \bibAnnoteFile{NoStop}{Chen11}%
\bibitem{Maier12}%
  \BibitemOpen
  \bibfield{author}{%
  \bibinfo {author} {\bibfnamefont{T.~A.}\ \bibnamefont{Maier}}, \bibinfo
  {author} {\bibfnamefont{P.~J.}\ \bibnamefont{Hirschfeld}},\ and\ \bibinfo
  {author} {\bibfnamefont{D.~J.}\ \bibnamefont{Scalapino}},\ }%
  \bibfield{journal}{%
  \bibinfo {journal} {Phys. Rev. B}\ }%
  \textbf{\bibinfo {volume} {86}},\ \bibinfo {pages} {094514} (\bibinfo {year}
  {2012})%
  \bibAnnoteFile{NoStop}{Maier12}%
\bibitem{Note1}%
  \BibitemOpen
  \bibinfo {note} {A simple valence counting of the contributing electrons
  together with the available bands below the Fermi level yields the formula
  $n=(12+x-8y)/(2-y)$ for the filling.}%
  \bibAnnoteFile{Stop}{Note1}%
\bibitem{Kuroki08}%
  \BibitemOpen
  \bibfield{author}{%
  \bibinfo {author} {\bibfnamefont{K.}~\bibnamefont{Kuroki}}, \emph{et~al.},\
  }%
  \bibfield{journal}{%
  \bibinfo {journal} {Phys. Rev. Lett.}\ }%
  \textbf{\bibinfo {volume} {101}},\ \bibinfo {pages} {087004} (\bibinfo {year}
  {2008})%
  \bibAnnoteFile{NoStop}{Kuroki08}%
\bibitem{a_kemper_10}%
  \BibitemOpen
  \bibfield{author}{%
  \bibinfo {author} {\bibfnamefont{A.~F.}\ \bibnamefont{Kemper}},
  \emph{et~al.},\ }%
  \bibfield{journal}{%
  \bibinfo {journal} {New J. Phys.}\ }%
  \textbf{\bibinfo {volume} {12}},\ \bibinfo {pages} {073030} (\bibinfo {year}
  {2010})%
  \bibAnnoteFile{NoStop}{a_kemper_10}%
\bibitem{Takimoto04}%
  \BibitemOpen
  \bibfield{author}{%
  \bibinfo {author} {\bibfnamefont{T.}~\bibnamefont{Takimoto}}, \bibinfo
  {author} {\bibfnamefont{T.}~\bibnamefont{Hotta}},\ and\ \bibinfo {author}
  {\bibfnamefont{K.}~\bibnamefont{Ueda}},\ }%
  \bibfield{journal}{%
  \bibinfo {journal} {Phys. Rev. B}\ }%
  \textbf{\bibinfo {volume} {69}},\ \bibinfo {pages} {104504} (\bibinfo {year}
  {2004})%
  \bibAnnoteFile{NoStop}{Takimoto04}%
\bibitem{s_graser_09}%
  \BibitemOpen
  \bibfield{author}{%
  \bibinfo {author} {\bibfnamefont{S.}~\bibnamefont{Graser}}, \bibinfo {author}
  {\bibfnamefont{T.~A.}\ \bibnamefont{Maier}}, \bibinfo {author}
  {\bibfnamefont{P.~J.}\ \bibnamefont{Hirschfeld}},\ and\ \bibinfo {author}
  {\bibfnamefont{D.~J.}\ \bibnamefont{Scalapino}},\ }%
  \bibfield{journal}{%
  \bibinfo {journal} {New J. Phys.}\ }%
  \textbf{\bibinfo {volume} {11}},\ \bibinfo {pages} {025016} (\bibinfo {year}
  {2009})%
  \bibAnnoteFile{NoStop}{s_graser_09}%
\bibitem{ABG}%
  \BibitemOpen
  \bibfield{author}{%
  \bibinfo {author} {\bibfnamefont{D.~F.}\ \bibnamefont{Agterberg}}, \bibinfo
  {author} {\bibfnamefont{V.}~\bibnamefont{Barzykin}},\ and\ \bibinfo {author}
  {\bibfnamefont{L.~P.}\ \bibnamefont{Gor'kov}},\ }%
  \bibfield{journal}{%
  \bibinfo {journal} {Phys. Rev. B}\ }%
  \textbf{\bibinfo {volume} {60}},\ \bibinfo {pages} {14868} (\bibinfo {year}
  {1999})%
  \bibAnnoteFile{NoStop}{ABG}%
\bibitem{Lee08}%
  \BibitemOpen
  \bibfield{author}{%
  \bibinfo {author} {\bibfnamefont{P.~A.}\ \bibnamefont{Lee}}\ and\ \bibinfo
  {author} {\bibfnamefont{X.-G.}\ \bibnamefont{Wen}},\ }%
  \bibfield{journal}{%
  \bibinfo {journal} {Phys. Rev. B}\ }%
  \textbf{\bibinfo {volume} {78}},\ \bibinfo {pages} {144517} (\bibinfo {year}
  {2008})%
  \bibAnnoteFile{NoStop}{Lee08}%
\bibitem{Hu13B}%
  \BibitemOpen
  \bibfield{author}{%
  \bibinfo {author} {\bibfnamefont{J.}~\bibnamefont{Hu}},\ }%
  \bibfield{journal}{%
  \bibinfo {journal} {Phys. Rev. X}\ }%
  \textbf{\bibinfo {volume} {3}},\ \bibinfo {pages} {031004} (\bibinfo {year}
  {2013})%
  \bibAnnoteFile{NoStop}{Hu13B}%
\bibitem{Hu13a}%
  \BibitemOpen
  \bibfield{author}{%
  \bibinfo {author} {\bibfnamefont{N.}~\bibnamefont{{Hao}}}\ and\ \bibinfo
  {author} {\bibfnamefont{J.}~\bibnamefont{{Hu}}},\ }%
  \bibfield{journal}{%
  \bibinfo {journal} {ArXiv e-prints}}%
   (\bibinfo {year} {2013}),\
  \Eprint{http://arxiv.org/abs/1305.5034}{arXiv:1305.5034 [cond-mat.supr-con]}%
  \bibAnnoteFile{NoStop}{Hu13a}%
\bibitem{Vafek13}%
  \BibitemOpen
  \bibfield{author}{%
  \bibinfo {author} {\bibfnamefont{V.}~\bibnamefont{{Cvetkovic}}}\ and\
  \bibinfo {author} {\bibfnamefont{O.}~\bibnamefont{{Vafek}}},\ }%
  \bibfield{journal}{%
  \bibinfo {journal} {ArXiv e-prints}}%
   (\bibinfo {year} {2013}),\
  \Eprint{http://arxiv.org/abs/1304.3723}{arXiv:1304.3723 [cond-mat.str-el]}%
  \bibAnnoteFile{NoStop}{Vafek13}%
\bibitem{Fischer13}%
  \BibitemOpen
  \bibfield{author}{%
  \bibinfo {author} {\bibfnamefont{M.~H.}\ \bibnamefont{Fischer}},\ }%
  \bibfield{journal}{%
  \bibinfo {journal} {New Journal of Physics}\ }%
  \textbf{\bibinfo {volume} {15}},\ \bibinfo {pages} {073006} (\bibinfo {year}
  {2013})%
  \bibAnnoteFile{NoStop}{Fischer13}%
\bibitem{Sorella13}%
  \BibitemOpen
  \bibfield{author}{%
  \bibinfo {author} {\bibfnamefont{M.}~\bibnamefont{{Casula}}}\ and\ \bibinfo
  {author} {\bibfnamefont{S.}~\bibnamefont{{Sorella}}},\ }%
  \bibfield{journal}{%
  \bibinfo {journal} {ArXiv e-prints}}%
   (\bibinfo {year} {2013}),\
  \Eprint{http://arxiv.org/abs/1302.4748}{arXiv:1302.4748 [cond-mat.supr-con]}%
  \bibAnnoteFile{NoStop}{Sorella13}%
\bibitem{Khodas12_PRB}%
  \BibitemOpen
  \bibfield{author}{%
  \bibinfo {author} {\bibfnamefont{M.}~\bibnamefont{Khodas}}\ and\ \bibinfo
  {author} {\bibfnamefont{A.~V.}\ \bibnamefont{Chubukov}},\ }%
  \bibfield{journal}{%
  \bibinfo {journal} {Phys. Rev. B}\ }%
  \textbf{\bibinfo {volume} {86}},\ \bibinfo {pages} {144519} (\bibinfo {year}
  {2012})%
  \bibAnnoteFile{NoStop}{Khodas12_PRB}%
\bibitem{Inosov11}%
  \BibitemOpen
  \bibfield{author}{%
  \bibinfo {author} {\bibfnamefont{J.~T.}\ \bibnamefont{Park}}, \emph{et~al.},\
  }%
  \bibfield{journal}{%
  \bibinfo {journal} {Phys. Rev. Lett.}\ }%
  \textbf{\bibinfo {volume} {107}},\ \bibinfo {pages} {177005} (\bibinfo {year}
  {2011})%
  \bibAnnoteFile{NoStop}{Inosov11}%
\bibitem{Inosov12}%
  \BibitemOpen
  \bibfield{author}{%
  \bibinfo {author} {\bibfnamefont{G.}~\bibnamefont{Friemel}}, \emph{et~al.},\
  }%
  \bibfield{journal}{%
  \bibinfo {journal} {Phys. Rev. B}\ }%
  \textbf{\bibinfo {volume} {85}},\ \bibinfo {pages} {140511} (\bibinfo {year}
  {2012})%
  \bibAnnoteFile{NoStop}{Inosov12}%
\bibitem{Maier09}%
  \BibitemOpen
  \bibfield{author}{%
  \bibinfo {author} {\bibfnamefont{T.~A.}\ \bibnamefont{Maier}}, \bibinfo
  {author} {\bibfnamefont{S.}~\bibnamefont{Graser}}, \bibinfo {author}
  {\bibfnamefont{D.~J.}\ \bibnamefont{Scalapino}},\ and\ \bibinfo {author}
  {\bibfnamefont{P.~J.}\ \bibnamefont{Hirschfeld}},\ }%
  \bibfield{journal}{%
  \bibinfo {journal} {Phys. Rev. B}\ }%
  \textbf{\bibinfo {volume} {79}},\ \bibinfo {pages} {224510} (\bibinfo {year}
  {2009})%
  \bibAnnoteFile{NoStop}{Maier09}%
\bibitem{Eschrig09}%
  \BibitemOpen
  \bibfield{author}{%
  \bibinfo {author} {\bibfnamefont{H.}~\bibnamefont{Eschrig}}\ and\ \bibinfo
  {author} {\bibfnamefont{K.}~\bibnamefont{Koepernik}},\ }%
  \bibfield{journal}{%
  \bibinfo {journal} {Phys. Rev. B}\ }%
  \textbf{\bibinfo {volume} {80}},\ \bibinfo {pages} {104503} (\bibinfo {year}
  {2009})%
  \bibAnnoteFile{NoStop}{Eschrig09}%
\bibitem{Friedel64}%
  \BibitemOpen
  \bibfield{author}{%
  \bibinfo {author} {\bibfnamefont{J.}~\bibnamefont{Friedel}}, \bibinfo
  {author} {\bibfnamefont{P.}~\bibnamefont{Lenglart}},\ and\ \bibinfo {author}
  {\bibfnamefont{G.}~\bibnamefont{Leman}},\ }%
  \bibfield{journal}{%
  \bibinfo {journal} {J. Phys. Chem. of Solids}\ }%
  \textbf{\bibinfo {volume} {25}},\ \bibinfo {pages} {781 } (\bibinfo {year}
  {1964})%
  \bibAnnoteFile{NoStop}{Friedel64}%
\bibitem{Xu12}%
  \BibitemOpen
  \bibfield{author}{%
  \bibinfo {author} {\bibfnamefont{M.}~\bibnamefont{Xu}}, \emph{et~al.},\ }%
  \bibfield{journal}{%
  \bibinfo {journal} {Phys. Rev. B}\ }%
  \textbf{\bibinfo {volume} {85}},\ \bibinfo {pages} {220504} (\bibinfo {year}
  {2012})%
  \bibAnnoteFile{NoStop}{Xu12}%
\bibitem{PBE}%
  \BibitemOpen
  \bibfield{author}{%
  \bibinfo {author} {\bibfnamefont{J.~P.}\ \bibnamefont{Perdew}}, \bibinfo
  {author} {\bibfnamefont{K.}~\bibnamefont{Burke}},\ and\ \bibinfo {author}
  {\bibfnamefont{M.}~\bibnamefont{Ernzerhof}},\ }%
  \bibfield{journal}{%
  \bibinfo {journal} {Phys. Rev. Lett.}\ }%
  \textbf{\bibinfo {volume} {77}},\ \bibinfo {pages} {3865} (\bibinfo {year}
  {1996})%
  \bibAnnoteFile{NoStop}{PBE}%
\bibitem{Guo10}%
  \BibitemOpen
  \bibfield{author}{%
  \bibinfo {author} {\bibfnamefont{J.}~\bibnamefont{Guo}}, \emph{et~al.},\ }%
  \bibfield{journal}{%
  \bibinfo {journal} {Phys. Rev. B}\ }%
  \textbf{\bibinfo {volume} {82}},\ \bibinfo {pages} {180520} (\bibinfo {year}
  {2010})%
  \bibAnnoteFile{NoStop}{Guo10}%
\bibitem{Shein11}%
  \BibitemOpen
  \bibfield{author}{%
  \bibinfo {author} {\bibfnamefont{I.}~\bibnamefont{Shein}}\ and\ \bibinfo
  {author} {\bibfnamefont{A.}~\bibnamefont{Ivanovskii}},\ }%
  \bibfield{journal}{%
  \bibinfo {journal} {Physics Letters A}\ }%
  \textbf{\bibinfo {volume} {375}},\ \bibinfo {pages} {1028 } (\bibinfo {year}
  {2011})%
  \bibAnnoteFile{NoStop}{Shein11}%
\bibitem{Cottenier}%
  \BibitemOpen
  \bibfield{author}{%
  \bibinfo {author} {\bibfnamefont{S.}~\bibnamefont{Cottenier}},\ }%
  \bibfield{journal}{%
  \bibinfo {journal} {Instituut voor Kern- en Stralingsfysica, K.U.Leuven,
  Belgium, ISBN 90-807215-1-4}}%
   (\bibinfo {year} {2002})%
  \bibAnnoteFile{NoStop}{Cottenier}%
\bibitem{Note2}%
  \BibitemOpen
  \bibinfo {note} {The spin-orbit coupling on the Se-atoms does not
  significantly influence the band energies close to the Fermi level. We do not
  consider it further although Se ($Z=34$) in general leads to stronger
  relativistic effects than Fe ($Z=26$).}%
  \bibAnnoteFile{Stop}{Note2}%
\end{thebibliography}
\end{document}